\documentclass[aps,pra,reprint,superscriptaddress]{revtex4-1}
\usepackage{amsmath}
\usepackage{graphicx}
\DeclareMathOperator*{\argmax}{argmax}
\usepackage{resizegather}

\usepackage{amssymb}
\usepackage{physics}
\usepackage{braket}
\usepackage{dsfont}
\usepackage{mathtools}
\usepackage{diagbox}
\usepackage{varwidth}
\usepackage{framed}
\usepackage{natbib}
\usepackage{soul,xcolor} 
\usepackage{bm}
\usepackage{textgreek}

\usepackage{epstopdf}
\usepackage{float}
\usepackage{caption}
\usepackage{tikz} 
\usetikzlibrary{decorations.pathmorphing,patterns}
\usetikzlibrary{backgrounds}
\usetikzlibrary{decorations.markings}
\tikzset{
    set arrow inside/.code={\pgfqkeys{/tikz/arrow inside}{#1}},
    set arrow inside={end/.initial=>, opt/.initial=},
    /pgf/decoration/Mark/.style={
        mark/.expanded=at position #1 with
        {
            \noexpand\arrow[\pgfkeysvalueof{/tikz/arrow inside/opt}]{\pgfkeysvalueof{/tikz/arrow inside/end}}
        }
    },
    arrow inside/.style 2 args={
        set arrow inside={#1},
        postaction={
            decorate,decoration={
                markings,Mark/.list={#2}
            }
        }
    },
}
\usetikzlibrary{shapes}
\definecolor{myred}{HTML}{FF0000} 
\definecolor{QRc}{HTML}{E64F40} 
\definecolor{nonQRc}{HTML}{3C49E3} 
\definecolor{eprsource}{rgb}{0.34, 0.81, 0.54}

\usepackage{xcolor}
\usepackage{mdframed}
\usepackage{multienum}

\makeatletter
\newcommand*{\centerfloat}{%
  \parindent \z@
  \leftskip \z@ \@plus 1fil \@minus \textwidth
  \rightskip\leftskip
  \parfillskip \z@skip}
\makeatother

\usepackage{todonotes}

\usepackage{hyperref}

\usetikzlibrary{shapes.geometric}
\usetikzlibrary{patterns}
\usetikzlibrary{positioning}
\usepackage[T1]{fontenc}
\usepackage{lmodern}
\usepackage{notoccite}
\usepackage{amssymb}

\tikzset{
    buffer/.style={
        draw,
        shape border rotate=180,
        regular polygon,
        regular polygon sides=3,
        fill=gray, fill opacity = 0.2,
        node distance=2cm,
        minimum height=4em
    }
}

\usepackage[noend]{algpseudocode}
\usepackage{algorithm}
\algnewcommand{\Initialize}[1]{%
  \State \textbf{Initialize:}
  \Statex \hspace*{\algorithmicindent}\parbox[t]{.8\linewidth}{\raggedright #1}
}

\makeatletter
\def\BState{\State\hskip-\ALG@thistlm}
\makeatother

\setstcolor{red} 

\begin{document}
\title{Inhomogeneous driving in quantum annealers can result in orders-of-magnitude improvements in performance} 
\author{Juan~I.~Adame}
\email{juan.adame@qcware.com}
\affiliation{QC Ware Corp., 125 University Ave., Suite 260, Palo Alto, CA 94301, USA,}
\author{Peter~L.~McMahon}
\email{pmcmahon@stanford.edu}
\affiliation{Ginzton Laboratory, Stanford University, Stanford, CA 94305, USA}
\affiliation{QC Ware Corp., 125 University Ave., Suite 260, Palo Alto, CA 94301, USA,}
\begin{abstract}
Quantum annealers are special-purpose quantum computers that primarily target solving Ising optimization problems. Theoretical work has predicted that the probability of a quantum annealer ending in a ground state can be dramatically improved if the spin driving terms, which play a crucial role in the functioning of a quantum annealer, have different strengths for different spins; that is, they are inhomogeneous. In this paper we describe a time-shift-based protocol for inhomogeneous driving and demonstrate, using an experimental quantum annealer, the performance of our protocol on a range of hard Ising problems that have been well-studied in the literature.  Compared to the homogeneous-driving case, we find that we are able to increase the probability of finding a ground state by up to $10^8 \times$ for some Weak-Strong-Cluster problem instances, and by up to $10^3 \times$ for more general spin-glass problem instances.  In addition to being of practical interest as a heuristic speedup method, inhomogeneous driving may also serve as a useful tool for investigations into the physics of experimental quantum annealers.
\end{abstract}
    \maketitle
\onecolumngrid
\section{Introduction}

Quantum annealing is a form of quantum computation that is primarily targeted at solving Ising combinatorial optimization problems \cite{kadowaki1998quantum,farhi2000quantum,santoro2006optimization,johnson2011quantum}.
In recent years, there has been great interest in finding whether or not an experimental quantum annealer (QA) can deliver a speedup over the best classical heuristic optimization methods \cite{ronnow2014defining, boixo2014evidence, denchev2016computational, hen2015probing, mandra2016strengths, albash2017evidence, mandra2017deceptive}.
Considerable effort has been put into understanding both what classes of problems might be most amenable to speedup on current experimental systems, as well as the design of modifications to current quantum annealing systems and protocols that may result in improved performance \cite{rams2016inhomogeneous,mohseni2018engineering,campo2013universal,nishimori2017exponential, susa2017relation,hormozi2017nonstoquastic,vinci2017non,susa2018exponential,gomez2018universal,venturelli2018reverse}.
The prospect of experimental quantum annealers delivering a speedup has resulted in a large volume of work exploring potential applications for future quantum annealers, ranging from particle physics \cite{mott2017solving},
to statistics \cite{lokhov2018optimal},
to bioinformatics \cite{li2018quantum}.

A quantum annealer operates \cite{kadowaki1998quantum}
by starting with strong quantum-fluctuation terms, called driving terms, that are slowly brought to zero by the end of the computation. Simultaneously, spin-spin couplings and external-field terms, which encode the problem to be solved, are increased from zero between the beginning and the end of the computation. Ideally the QA will end in a ground state of the encoded optimization problem. In practice, the probability of a QA finding a ground state at the end of any particular annealing run is far less than 100\% \cite{ronnow2014defining} --
probabilities of $10^{-7}$, or even smaller, are routinely observed for many problems on current experimental systems. A longstanding goal of the quantum-annealing community is to discover principles and methods that result in the probability of finding a ground state being maximized. Theoretical work has predicted that dramatic improvements in the success probability can be achieved if the driving terms are applied with different strengths to different spins; that is, they are inhomogeneous \cite{rams2016inhomogeneous,mohseni2018engineering,campo2013universal,susa2018exponential,gomez2018universal}.
Recently it has become possible to experimentally test a restricted form of inhomogeneous driving, in which one does not have arbitrary control over the driving terms, but one can delay or advance the driving schedule on a qubit-by-qubit basis \cite{andriyash2016factoring,lanting2017experimental}.
There are an exponential (in number of qubits, $N$) number of possible choices for advancing or delaying the qubits's driving schedules, which provides scope for the investigation of a wide variety of strategies for using so-called {\it anneal offsets} (AO) to improve the performance of quantum annealers. In this paper we describe one particular strategy, which is distinct from the strategies reported in Refs. \cite{andriyash2016factoring,lanting2017experimental},
and we present results from experiments performed on the D-Wave 2000Q (DW2000Q) QA \cite{DW2000Q} hosted at NASA Ames in which we show how this AO strategy improves the performance of the annealer across a range of benchmark problems that have previously been studied in the quantum-annealing literature \cite{ronnow2014defining,boixo2016computational,denchev2016computational,mandra2016strengths, mishra2018finite}.

We now introduce more formally the concept of anneal offsets, and the strategy that we investigate in this paper.

The canonical form of quantum annealing, which involves homogeneous driving terms, is described by the following time-dependent Hamiltonian for $N$ qubits:
\begin{equation}\label{Uniform QA}
H(s)=A(s)H_\textrm{D}+B(s)H_{\rm{P}},\quad H_\textrm{D}=\sum_{i=1}^{N}\sigma_{x}^{(i)},\quad H_{\rm{P}}=\sum_{i=1}^{N}h_{i}\sigma_{z}^{(i)}+\sum_{i=1}^{N}\sum_{j=1}^{i-1}J_{ij}\sigma_{z}^{(i)}\otimes\sigma_{z}^{(j)},
\end{equation}
where $s\in[0,1]$ is the normalized time parameter ($s=0$ is the start of the computation, and $s=1$ is the end), and superscripts $(i)$ and $(j)$ are qubit indices.  The Hamiltonian $H_{\rm{P}}$ is the so-called {\emph{problem Hamiltonian}}, and its ground states encode the solutions to the classical energy-minimization problem of the Ising model, $\text{argmin}_{\bm{x}\in\{-1,+1\}^{N}} \sum_{i=1}^{N}h_{i}x_i+\sum_{i=1}^{N}\sum_{j=1}^{i-1}J_{ij}x_{i}x_{j}$, where $\bm{x}\coloneqq(x_1,\dots,x_N)$. The time-dependent terms in Eq.~\eqref{Uniform QA} are the schedules: $A(s)$ controls how the driving terms, given in $H_\textrm{D}$, are turned off between time $s=0$ and $s=1$, and $B(s)$ controls how the problem-Hamiltonian terms are turned on between time $s=0$ and time $s=1$. For this form of quantum annealing, the driving is homogeneous: the driving term for every qubit is turned off using the same schedule ($A(s)$).

The D-Wave 2000Q quantum annealer, when operating with homogeneous driving, has both schedules $A(s)$ and $B(s)$ controlled by a single time-dependent signal $c(s)$. Inhomogeneous driving is made possible by allowing this signal $c(s)$ to be qubit-dependent in a specific way \cite{andriyash2016factoring,lanting2017experimental}: for each qubit $i$, the signal defining its annealing schedules can be perturbed by an offset $\delta_{i}$, which results in its driving-term schedule and problem-Hamiltonian-terms schedule being modified. Formally there are now $N$ signals $c_{i}(s)$, which are set to $c_{i}(s) \coloneqq c(s)+\delta_{i}$, and these result in qubit-independent schedules $A_i(s)\coloneqq A(s,\delta_i)$ and $B_i(s)\coloneqq B(s,\delta_i)$.  Illustrations showing the relationships between the normalized annealing time $s$, the signals $c_{i}(s)$, and the schedules $A(s,\delta_i)$ and $B(s,\delta_i)$, are given in Figure~A\ref{fig:schedules}. We can define a vector specifying the anneal offsets for each qubit, $\bm{\delta}=(\delta_1,\dots,\delta_N)$, and the homogeneous-driving quantum annealing Hamiltonian in Eq. \eqref{Uniform QA} is modified to become the following inhomogeneous-driving Hamiltonian:
\begin{equation}\label{AO QA}
H_{\bm{\delta}}(s)=\sum_{i=1}^{N}A_{i}(s)\sigma_{x}^{(i)}+\sum_{i=1}^{N}B_{i}(s)h_{i}\sigma_{z}^{(i)}+\sum_{i=1}^{N}\sum_{j=1}^{i-1}\sqrt{B_{i}(s)B_{j}(s)}J_{ij}\sigma_{z}^{(i)}\otimes\sigma_{z}^{(j)}.
\end{equation}
The driving terms $\sigma_{x}^{(i)}$ now have per-qubit schedules $A_{i}(s)$. Figure~\ref{fig:intro_fig} shows the annealing schedules for a problem with two clusters of qubits, where in one case homogeneous driving is used (Fig.~\ref{fig:intro_fig}a), and in the other, inhomogeneous driving -- where the schedules for the qubits in the second cluster are delayed -- is used (Fig.~\ref{fig:intro_fig}b).

How should the anneal offsets $\bm{\delta}$ be chosen to increase (ideally maximally) the probability of successfully finding a ground state? In the Appendix we show that the problem of finding optimal offsets can be viewed as the problem of optimizing a non-linear, high-dimensional functional that itself depends on the solution of the encoded minimization problem. Thus a formal approach seems impractical. Instead we may resort to heuristic approaches. Two prior experimental studies have outlined two different approaches for different classes of problem instances. Andriyash et al. \cite{andriyash2016factoring} report a strategy for problems that are embedded in a physical QA graph such that each logical variable in the original problem is represented by a chain of physical qubits, which occurs on the D-Wave QA when the problem to be solved is not a subgraph of the QA's Chimera physical hardware graph. Their strategy is to apply to each qubit a delay that increases monotonically with the length of the chain that the qubit forms part of. They tested their strategy on integer-factoring problems embedded in the Chimera graph, and reported improvements of up to $\sim 10^3 \times$. Lanting et al. \cite{lanting2017experimental} present an iterative approach to choosing the anneal offsets based on discovering the \emph{floppiness} of each qubit (which they relate to the classical notion of a floppy spin---a spin that does not change the system energy if it is flipped), and setting the offsets based on floppiness. They tested their strategy on a class of crafted instances of size up to $N=24$ qubits; larger problem instances were not explored.

\begin{figure}[!htb]
\includegraphics[width=0.8\textwidth,keepaspectratio]{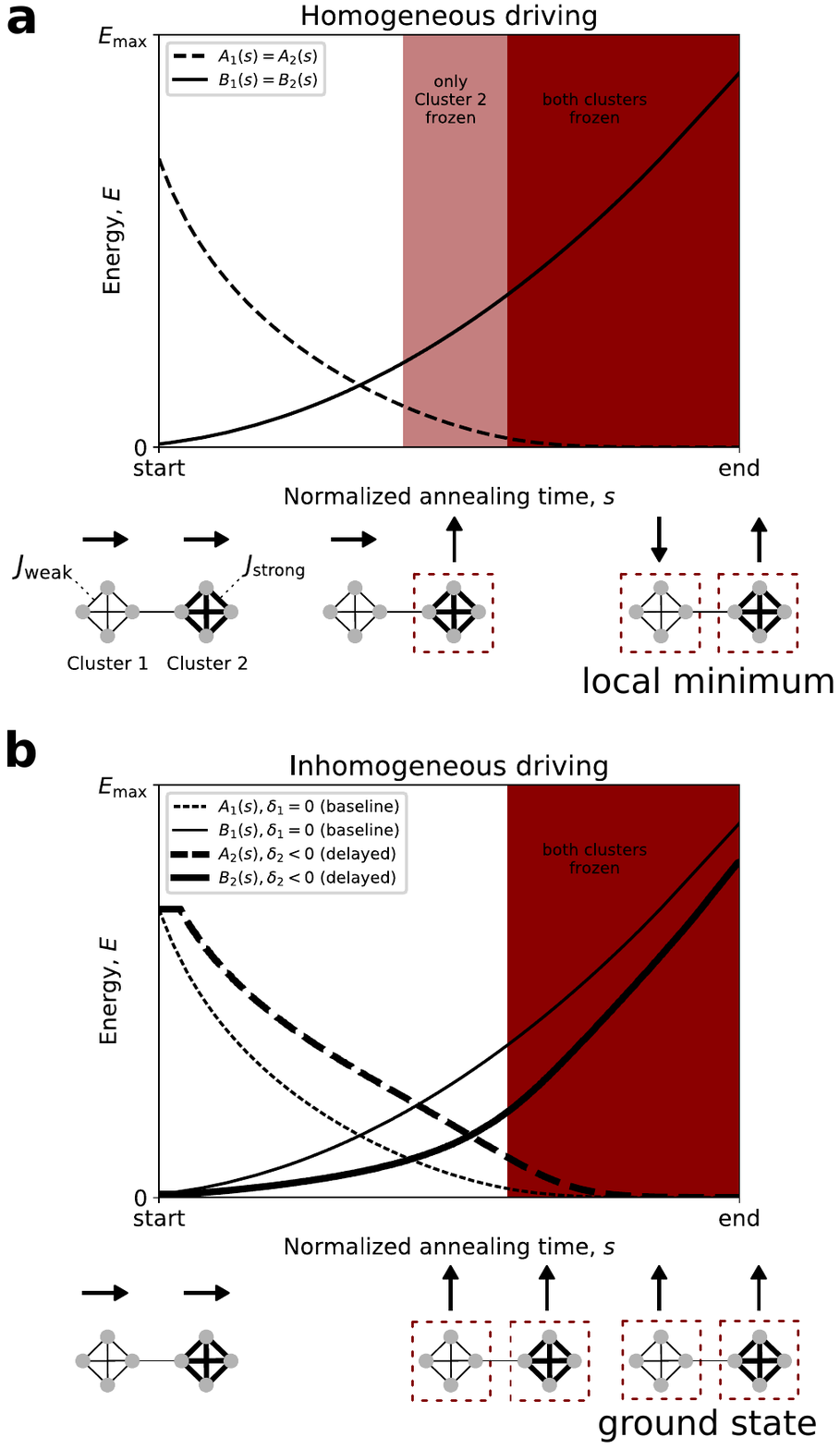}\label{fig:intro_fig}
\begin{flushleft}
\caption{}\label{fig:intro_fig}\footnotesize{{\bf{Fig.~\ref{fig:intro_fig}}}  ``Cartoon'' depiction of freeze-out for a toy problem. {\bf{a}} Under homogeneous driving, the more strongly coupled Cluster 2 freezes independently before the more weakly-coupled Cluster 1, causing the QA to end up in a local minimum.  {\bf{b}}  By delaying the annealing schedules of Cluster 2, with respect to the annealing schedules of Cluster 1, the freeze-out times of Cluster 1 and Cluster 2 are synchronized and the QA ends up in a ground state.}
\end{flushleft}
\end{figure}

One prominent hypothesis for why quantum annealers may fail to find a ground state is the phenomenon of freeze-out, in which certain qubits are thought to become frozen long before the end of the computation \cite{boixo2016computational,dickson2013thermally,mohseni2018constructing,marshall2018power}.  A hypothesized intuition for how anneal offsets can improve success probabilities is by delaying the reduction of the driving terms for qubits that are prone to early freeze-out. The aforementioned Ref. \cite{andriyash2016factoring} applied this intuition in the context of problems with qubit chains, where qubits in long chains are claimed to freeze earlier than those in short chains. In this paper we experimentally explore an AO strategy that is based primarily on two hypotheses: early freeze-out of a qubit can be mitigated by delaying annealing schedules for that qubit, and qubits that are more strongly coupled to the rest of the system freeze out earlier than those that are only weakly coupled \cite{test1}.

We now give a precise definition of the strategy that we experimentally tested. One key component of the strategy is that we quantify how strongly coupled a spin is to the rest of the system via an \emph{effective field}.  Let $j_1,\dots,j_{N_i}$ denote an enumeration of the spins that are coupled to spin $i$, and let $s_{j_1},\dots,s_{j_{N_i}}\in\{+1,-1\}$ be some configuration of these spins.  We denote the effective field on spin $i$, as a function of the value of the spins neighboring spin $i$,
by\clearpage\noindent$\mathcal{F}_i(s_{j_1},\dots,s_{j_{N_i}})$, where this is defined as
\begin{equation}
\mathcal{F}_i(s_{j_1},\dots,s_{j_{N_i}}) \coloneqq  h_i + \sum_{j=j_1,\dots,j_{N_i}} J_{ij} s_j.
\end{equation}
We will primarily be interested specifically in the absolute value of this quantity, $|\mathcal{F}_i(s_{j_1},\dots,s_{j_{N_i}}))|$, and in particular, the average of this over all possible neighboring spin values:
\begin{equation}
\label{F_i}
\overline{|\mathcal{F}_i|} \coloneqq \frac{1}{2^{N_i}}\sum_{s_{j_1},\dots,s_{j_{N_i}}\in\{+1,-1\}}|\mathcal{F}_i(s_{j_1},\dots,s_{j_{N_i}})|.
\end{equation}
Next, we normalize these averages to obtain values in the interval $[0,1]$:
\begin{equation}\label{ratios}
r_i\coloneqq\frac{\overline{|\mathcal{F}_i|}}{\max\limits_{k\in\{1,\dots,N\}}\overline{|\mathcal{F}_k|}}.
\end{equation}
A function that delays qubits based on the magnitude of their average effective fields is $\delta_i({r_i})\coloneqq|\delta|_{\max}(1-2r_i)$, where $|\delta|_{\max}>0$ is the maximum magnitude of offset that is applied on any qubit according to this method, and is a parameter that we are free to choose.  One detail we need to consider is that different offset ranges are available for different qubits, and it can be the case that this function assigns an offset to a qubit that the hardware cannot physically realize. If $\delta_i^{\max}$ is the maximum offset value that can be applied on qubit $i$ allowed by the hardware, and $\delta_i^{\min}$ is the minimum such offset value, we can ensure that $\delta_i({r_i}) \in [\delta_i^{\min},\delta_i^{\max}]$ by using the following prescription, which is the formal definition of the strategy that we explore in this paper: 
\begin{equation}\label{algorithm}
\delta_i({r_i})\coloneqq\left\{
\begin{array}{ll}
\min \{|\delta|_{\max}(1-2r_i), \delta_i^{\max}\}\qquad&{\rm{if}}\quad r_i\geq \frac{1}{2},\\
\max \{|\delta|_{\max}(1-2r_i), \delta_i^{\min}\}\qquad&{\rm{if}}\quad r_i<\frac{1}{2}.
\end{array}
\right.
\end{equation}

We note that the averages in Eq. \eqref{F_i} can be computed without difficulty for any graph that has low maximum degree, such as that of the DW2000Q (which has a maximum degree of 6, so only at most $2^6$ spin configurations have to be enumerated per qubit to calculate $\overline{|\mathcal{F}_i|}$). The computation defined in Eq.~\eqref{F_i} is intractable when the underlying QA hardware graph has vertices with high degree. However, in the Appendix, we present a modified version of the heuristic that is tractable even for fully connected graphs. Since in this paper we only work with the DW2000Q, we can directly perform the computations specified by Eq. \eqref{F_i}.

We experimentally tested the strategy in Eq.~(6) on several different classes of problems by comparing the performance of the DW2000Q using baseline settings (no anneal offsets) against the performance of the DW2000Q using anneal offsets with various choices of $|\delta|_{\max}$. \textit{Success} of a single annealing run is defined as finding a ground state of the given problem instance at the end of the anneal. Success probabilities were estimated by taking the number of observed successes, and dividing by the number of annealing runs performed to observe this number of successes.  From these observed success probabilities, $p$, the corresponding time-to-solution (TTS) with desired probability $p_{\rm{d}}$ (which we chose to be $99\%$), was computed using the formula $\textrm{TTS} \coloneqq t_{\rm{ann}}\log(1-p_{\rm{d}})/\log(1-p)$, where $t_{\rm{ann}}$ is the annealing time used in each run.  We used the default value $t_{\rm{ann}}=20~$\textmu s for all problem classes except for the Alternating-Sectors-Chain problems, for which we used $t_{\rm{ann}}=5~$\textmu s to allow for direct comparison with results from a recent baseline experimental study \cite{mishra2018finite}. Furthermore, all problem classes were defined on the Chimera graph (native to the DW2000Q), except for the Alternating Sectors Chain, which is a 1-D chain that embeds directly into the native Chimera graph.

\section{Results}
\subsection*{Uniform-Range-$k$-Disorder (UR$k$D) problems}

\begin{figure}[!htb]
\includegraphics[width=0.9\textwidth, keepaspectratio]{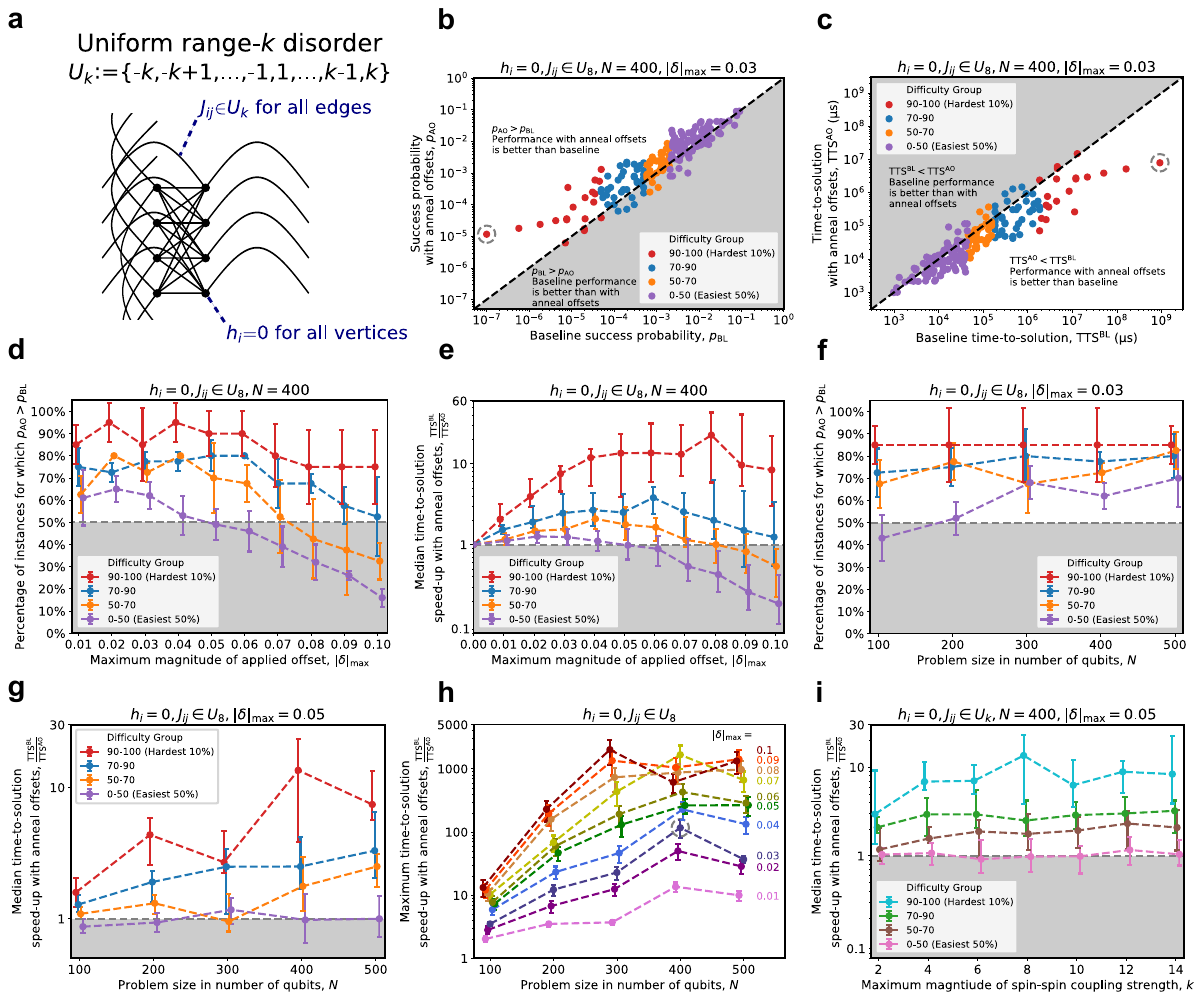}
\caption{}
\label{fig:Uk}
\begin{flushleft}
\footnotesize{\noindent{\bf{Fig.~\ref{fig:Uk}}}  Results for the Uniform-Range-$k$ Disorder (UR$k$D) problem class.
{\bf{a}} A zoom-in on one of the Chimera cells in a UR$k$D problem instance.  The coupler values $J_{ij}$ are chosen at random, with uniform probability, from the set $U_k\coloneqq\{-k,-k+1,\dots,-1,1,\dots,k-1,k\}$.  Note that $0\notin U_k$.  The local fields $h_v$ are all set to 0 (i.e, $h_v=0$ for all $v\in V$).  {\bf{b}}  Instance-by-instance comparison of the observed success probability when using anneal offsets, $p_{\rm{AO}}$, versus the observed success probability with the baseline schedule, $p_{\rm{BL}}$. {\bf{c}}  The corresponding times-to-solution, calculated directly from the success probabilities in {\bf{b}}; $\rm{TTS}^{\rm{AO}}$ denotes the time-to-solution when using anneal offsets, and $\rm{TTS}^{\rm{BL}}$ denotes the time-to-solution with the baseline schedule.  The instance for which the maximum improvement was observed is emphasized by a grey, dashed circle.  Instances in the white zone correspond to those for which the schedule with anneal offsets resulted in better performance, whereas those in the grey zone correspond to those for which the baseline schedule resulted in better performance.  The color indicates the relative difficulty of the instance as measured by the performance with the baseline schedule.  {\bf{d}} Percentage of instances for which an improvement in the success probability when using anneal offsets was observed, versus the maximum magnitude of the applied offsets, $|\delta|_{\max}$. {\bf{e}} Median speed-up observed when using anneal offsets versus $|\delta|_{\max}$.  {\bf{f}} Percentage of instances for which an improvement in the success probability when using anneal offsets was observed, versus $N$.  {\bf{g}} Median TTS speed-up observed when using anneal offsets versus $N$.  {\bf{h}} The maximum TTS speed-up observed when using anneal offsets versus $N$, for all the different values of $|\delta|_{\max}$ used.  {\bf{i}} Median TTS speed-up observed when using anneal offsets versus the spin-spin coupling range, $k$.  Note that $k$ is held fixed (at $k=8$) in the rest of the figure.}
\end{flushleft}
\end{figure}

The Uniform-Range-$k$-Disorder (UR$k$D) class of problems is defined (Fig.~\ref{fig:Uk}a) on the Chimera graph as those for which $h_i=0$ for all $i\in\{1,\dots,N\}$, and each $J_{ij}$ is chosen at random, with uniform probability, from the $2k$ discrete values in the set $U_k\coloneqq\{-k,-k+1,\dots,-1,1,\dots,k-1,k\}$.  
These problems have previously been studied in the context of quantum annealing in Refs. \cite{zhu2016best, ronnow2014defining,boixo2014evidence}.  The generic nature of this problem class makes it a good candidate for getting a sense for how useful the heuristic strategy in Eq.~\eqref{algorithm} might be on optimization problems that arise in a variety of application areas. We generated random instances of problems in the UR$k$D class, for various graph sizes $N$ and coupling ranges $k$,
and measured the success probability of finding a ground state for each instance, both with baseline DW2000Q operation, and when using anneal offsets as prescribed by the heuristic. Our experimental results, summarized in Fig.~\ref{fig:Uk}, show that anneal offsets chosen using the heuristic typically improve performance as measured by several different metrics.

Figure~\ref{fig:Uk}b shows a scatter plot of the observed success probability when using anneal offsets, $p_{\rm{AO}}$, versus the observed success probability with the baseline schedule, $p_{\rm{BL}}$, for 200 randomly generated instances of problem size $N=400$ and maximum magnitude of spin-spin coupling $k=8$, using a maximum magnitude of applied offset $|\delta|_{\max}=0.03$.  Figure~\ref{fig:Uk}c shows the corresponding times-to-solution, calculated directly from the success probabilities in Fig.~\ref{fig:Uk}b; $\rm{TTS}^{\rm{AO}}$ denotes the time-to-solution when using anneal offsets, and $\rm{TTS}^{\rm{BL}}$ denotes the time-to-solution with the baseline schedule. In Fig.~\ref{fig:Uk}b,c, we have colored the instances by the difficulty with which the DW2000Q solves them using baseline settings. In particular, we have divided the instances into four difficulty groups, which we formally defined using ranges of the percentile-rank calculated for each instance based on its baseline success probability $p_\textrm{BL}$.  The grouping of instances by difficulty occurs in subsequent panels too, and in subsequent figures, and has been performed the same way throughout.  We note that the more difficult instances are both more likely to benefit from the use of anneal offsets, as well as more likely to benefit to a larger degree, relative to the easier instances in this problem class.  One can see this more clearly in Fig.~\ref{fig:Uk}d,e, which show the percentage of instances for which using anneal offsets resulted in improved performance compared to baseline, and the median time-to-solution ratio (i.e., the median of $\rm{TTS}^{\rm{BL}}/\rm{TTS}^{\rm{AO}}$), respectively, both as functions of $|\delta|_{\max}$.

It is natural to ask what value of $|\delta|_{\max}$ results in the ``best'' performance.  An answer to this question is not straightforward, and Fig.~\ref{fig:Uk}d,e give some insight into the trade-offs encountered through various choices of $|\delta|_{\max}$.  To start, determining the best choice of $|\delta|_{\max}$ depends on the performance metric used.  
Another fact that further complicates this question is that different instances will be affected differently for the same value of $|\delta|_{\max}$. 
For this problem class, the primary trade-offs to be balanced are that smaller values of $|\delta|_{\max}$ will generally result in increased performance over a larger percentage of the instances, but smaller median time-to-solution speed-ups for the more difficult instances, relative to larger values of $|\delta|_{\max}$.  For a clearer picture of these trade-offs, see the scatter plots in Fig.~A\ref{fig:Uk_supp}.

Another question that arises naturally is how the performance of the heuristic depends on problem size.  To that end, Fig.~\ref{fig:Uk}f shows the percentage of instances for which the use of anneal offsets resulted in improved performance for different problem sizes, $N$,
with $|\delta|_{\max}$ fixed at a value chosen based on Fig.~\ref{fig:Uk}d such that performance averaged over all instances is improved ($|\delta|_{\max}=0.03$).  While the change in this metric as a function of $N$ for each difficulty group is different, the overall impression is that as $N$ increases, the benefit obtained from using anneal offsets over baseline is either constant or increasing, depending on the difficulty group.  Furthermore, across all problem sizes there is a tendency for performance to be improved on a larger percentage of the more difficult instances, relative to the easier instances.  Fig.~\ref{fig:Uk}g shows the median time-to-solution ratio for various problem sizes, with $|\delta|_{\max}$ fixed at a value chosen based on Fig.~\ref{fig:Uk}e such that performance averaged over all instances is improved ($|\delta|_{\max}=0.05$).  While the behavior is again dependent on which difficulty group is being considered, in general there appears to be a tendency for the median time-to-solution ratio to increase slightly with problem size
for the hardest $50\%$ of instances, whereas for the easiest $50\%$ it appears to be constant.  Figure~\ref{fig:Uk}h shows the maximum observed speed-up (i.e., the maximum of $\rm{TTS}^{\rm{BL}}/\rm{TTS}^{\rm{AO}}$) for various problem sizes, for all values of $|\delta|_{\max}$ tested.  In general, there appears to be a tendency for the maximum speed-up to increase with increasing $N$, up to $N=400$, at which point the maximum speed-up plateaus.  The largest speed-up observed across all instances was just over $2000\times$ ($N=300$, $|\delta|_{\max}=0.1$).  We note that higher values of $|\delta|_{\max}$ tend to be more likely to result in the maximum speed-up, compared to the smaller values of $|\delta|_{\max}$.

The results discussed thus far have been for instances with $k=8$, where $k$ is the maximum magnitude of the spin-spin coupling values. We also performed experiments for a range of different $k$ values, and found the results to be broadly similar (Fig.~\ref{fig:Uk}i), suggesting that the use of anneal offsets provides benefit generally for UR$k$D instances, at least when the $U_k$ values are well within the precision limits of the QA ($2 \leq k \leq 14$ tested here).

\begin{figure*}[!ht]
\caption{}\label{fig:Uk_hybrid}
  \includegraphics[width=1\textwidth, keepaspectratio]{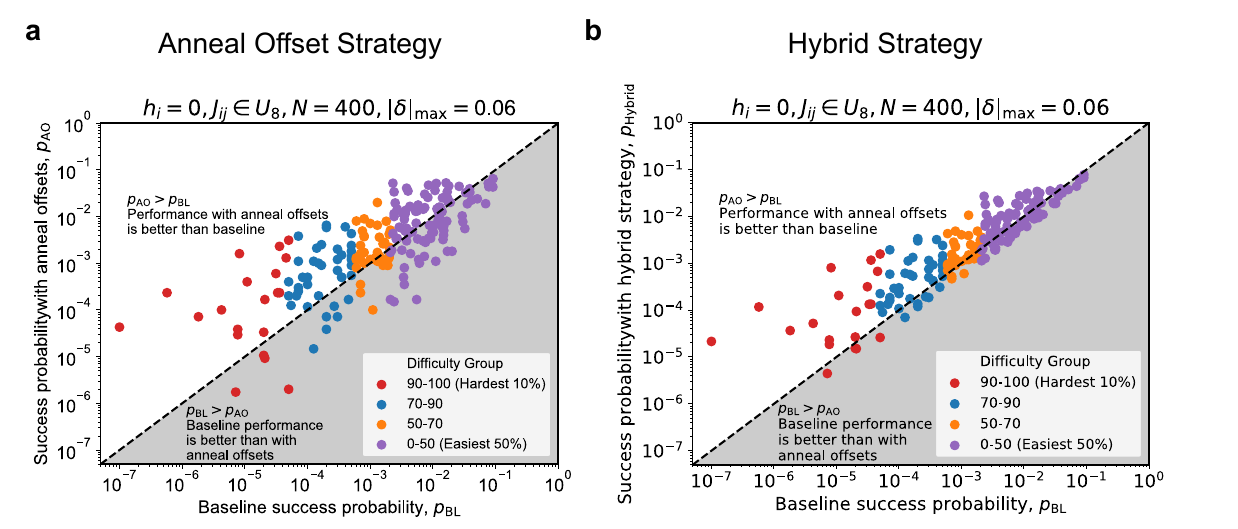}
  \begin{flushleft}
  \footnotesize{\noindent{\bf{Fig.~3}}.  Results for the Uniform-Range-$k$-Disorder (UR$k$D) problem class using the anneal-offset strategy and the hybrid strategy.
{\bf{a}} Instance-by-instance comparison of the observed success probability when using anneal offsets, $p_{\rm{AO}}$, with $|\delta|_{\max}=0.06$, versus the observed success probability with the baseline schedule, $p_{\rm{BL}}$. {\bf{b}} The corresponding success probability when using the hybrid strategy, $p_{\rm{Hybrid}}$, calculated directly from the data in {\bf{a}}, versus the observed success probability with the baseline schedule, $p_{\rm{BL}}$.}
  \end{flushleft}
\end{figure*}

\subsection*{A hybrid strategy for practical use of anneal offsets}

As we saw in the previous section, the impact of using AO on performance will in general be different for different problem instances, even when the same value of $|\delta|_{\max}$ is used.  In fact, for all values of $|\delta|_{\max}$ tried, there were always both instances that benefited from the use of AO as well instances for which using AO was detrimental to performance.  From the point of view of using anneal offsets in practice, this is a concern, since it is not known a priori whether it would be beneficial or not to use AO for any particular instance.  In this section we present a simple hybrid strategy that eliminates the risk that using AO may result in much worse performance. In particular, this hybrid strategy has the property that the strategy's time-to-solution (TTS) is provably at worst two times longer than the baseline TTS, but retains the ability to capture most of the benefit that using AO can bring.

The strategy consists of alternating between calls to the DW2000Q with baseline settings, and calls using anneal offsets.  Suppose for some problem instance the success probability using the baseline settings is $p_{\rm{BL}}$, and the success probability using a given anneal-offsets strategy is $p_{\rm{AO}}$.  Using the alternating strategy described above, the cumulative time-to-solution, which we denote ${\rm{TTS}}^{\rm{Hybrid}}$, is given by \begin{equation}{\rm{TTS}}^{\rm{Hybrid}}=2t_{\rm{ann}}\frac{\log(1-p_d)}{\log[(1-p_{\rm{BL}})(1-p_{\rm{AO}})]}.
\end{equation}
Because $p_{\rm{AO}}\geq0$, this implies the following upper bound:
\begin{equation}{\rm{TTS}}^{\rm{Hybrid}}\leq2t_{\rm{ann}}\frac{\log(1-p_d)}{\log(1-p_{\rm{BL}})}=2{\rm{TTS}}^{\rm{BL}}.
\end{equation}

Figure~3 shows a comparison of the standard AO strategy with this hybrid strategy for the same instances used in Fig.~2b.  More specifically, Fig.~3a shows a scatter plot of the observed success probability when using anneal offsets, $p_{\rm{AO}}$, versus the observed success probability with the baseline schedule, $p_{\rm{BL}}$, when $|\delta|_{\max}=0.06$ (in contrast to the value $|\delta|_{\max}=0.03$ used in Fig.~2b).  Note that, compared to Fig.~2b, Fig.~3a shows that in general there are greater improvements in success probability for harder instances, but also greater reductions in success probability for easier instances. Fig.~3b shows the results from applying the hybrid strategy to the same instances. The performance reductions are bounded in the hybrid strategy, while the performance improvements for harder instances are still observed.

When using anneal offsets, one may wish to limit the reductions in performance that can occur for easy instances by choosing a smaller value of $\left|\delta\right|_{\max}$. Unfortunately choosing a smaller value of $\left|\delta\right|_{\max}$ also reduces the performance benefit one obtains on difficult instances. By using the hybrid strategy one can use a higher value of $\left|\delta\right|_{\max}$ and obtain the bulk of the performance improvements from doing so while simultaneously limiting the downside for instances that aren't solved more easily using anneal offsets.

\subsection*{Alternating-Sectors-Chain (ASC) problems}

\begin{figure}[!h]
\includegraphics[width=\textwidth, keepaspectratio]{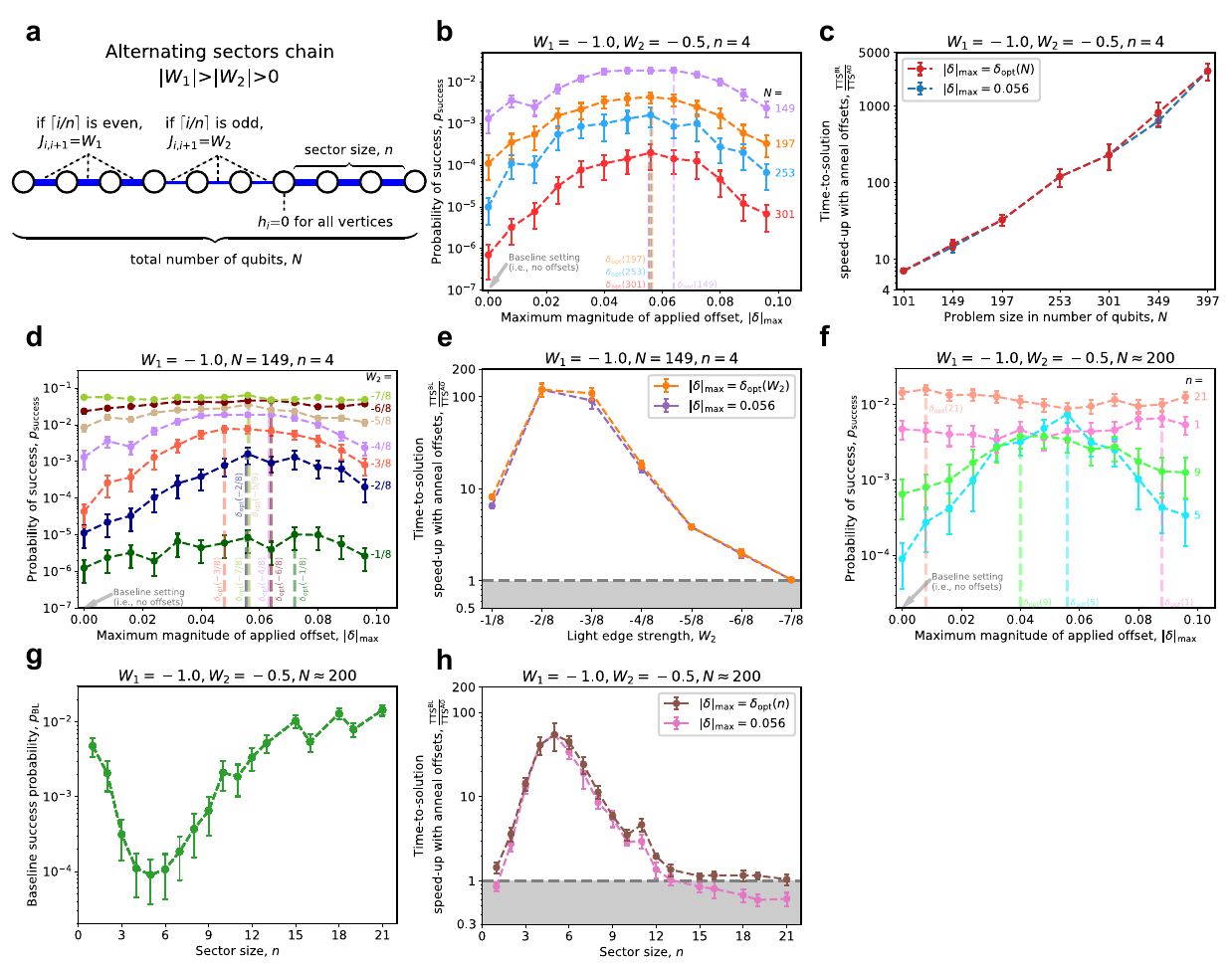}
\begin{flushleft}
\caption{}\label{fig:ASC}\noindent{\bf{Fig.~\ref{fig:ASC}}}  Results for the Alternating-Sectors-Chain (ASC) problem class.  {\bf{a}}  Schematic of an ASC problem instance.   {\bf{b}}  Success probability, $p_{\rm{success}}$, versus the maximum magnitude of the applied offsets, $|\delta|_{\max}$, for ASC problem instances of various problem sizes, $N$.  Recall that the case $|\delta|_{\max}=0$ is equivalent to using the baseline schedule.  The vertical dashed lines indicate the values of $|\delta|_{\max}$ at which the success probability is maximized for each $N$.  {\bf{c}}  Scaling of the time-to-solution (TTS) speed-up using anneal offsets versus $N$.  {\bf{d}}  $p_{\rm{success}}$ versus $|\delta|_{\max}$ for various light edge values, $W_2$.  The vertical dashed lines indicate the values of $|\delta|_{\max}$ at which the success probability is maximized for each $W_2$.  {\bf{e}}  TTS speed-up using anneal offsets versus $W_2$.  {\bf{f}}  $p_{\rm{success}}$ versus $|\delta|_{\max}$ for various sector sizes, $n$.  The vertical dashed lines indicate the values of $|\delta|_{\max}$ at which the success probability is maximized for each $n$.  {\bf{g}}  Baseline success probability, $p_{\rm{BL}}$, versus $n$.  {\bf{h}}  TTS speed-up using anneal offsets vs $n$.
\end{flushleft}
\end{figure}

The Alternating Sectors Chain (ASC) class of problems has been studied in the context of quantum annealing and adiabatic quantum computation in Refs. \cite{reichardt2004quantum,mishra2018finite}.  An Alternating Sectors Chain is a 1-dimensional (1-D) chain of $N$ spins divided into equally sized sectors of length $n$; sectors alternate between having so-called `heavy' ferromagnetic spin-spin couplings, $W_1$, and `light' ferromagnetic spin-spin couplings, $W_2$, where $|W_1|>|W_2|>0$.  Formally, if the spins are indexed by $\{1,\dots,N\}$, then ASC problems are given by $h_{i'}=0$ for all $i'\in\{1,\dots,N\}$, and for all $i\in\{1,\dots,N-1\}$ $J_{i,i+1}=W_1$ if $\lceil i/n\rceil$ is odd, and $J_{i,i+1}=W_2$ otherwise.  Furthermore, there is a technical restriction that there be $b+1$ heavy sectors and $b$ light sectors, which results in a limitation on the possible combinations of $N$ and $n$.  Figure~\ref{fig:ASC}A shows an example of an ASC instance with $N=10,n=3$.  Because all the couplings are ferromagnetic, the problem is trivial to solve: the two degenerate ground states are the fully aligned states, with all spins pointing either up or down.  Nevertheless, it is known that this problem exhibits an exponentially small gap in the sector size $n$ \cite{reichardt2004quantum}, which implies an exponential computation time in the AQC framework.  Recently, however, Ref. \cite{mishra2018finite} showed experimentally that the performance of a quantum annealer solving ASC problems differed substantially from what one would expect based purely on the scaling of the minimum gap.

We briefly paraphrase the intuitive argument of Ref.~\cite{mishra2018finite} for why quantum annealing fails to efficiently solve this problem.  For $N\gg1$ and $n\gg1$, any given sector approximates a 1-D ferromagnetic chain with all-equal couplings.  Such a chain encounters a quantum phase transition separating the ordered phase from the disordered phase when $A(s)=B(s)J_{i,i+1}$ \cite{sachdev2011quantum}  Therefore, the heavy sectors order independently before the light sectors during the anneal.  Since the transverse field generates only local spin flips, quantum annealing is likely to get stuck in a local minimum (with domain walls at the boundaries between heavy and light sectors) unless the annealing time is scaled exponentially with $n$.

The intuitive argument above suggests the following remedy via the use of anneal offsets.  Let $s_1$ be defined as the value such that  $A(s_1)=B(s_1)W_1$, and let $s_2$ be the value such that $A(s_2)=B(s_2)W_2$.  In other words, $s_1$ is the normalized time when the heavy sectors order, and $s_2$ is the normalized time when the light sectors order, under homogeneous driving. Let $\delta_1$ be an offset such that $A(s_2,\delta_1)=B(s_2,\delta_1)W_1$.  Applying this offset to the heavy sectors, one can make it so that both the light and heavy sectors order at the same time, $s_2$.  If it is the case that the sectors independently ordering at different times makes the problem more difficult to solve via quantum annealing, then we should see an increase in the success probability by applying this offset.  It turns out that anneal offsets calculated according to the prescription above are very similar to the anneal offsets that the heuristic in Eq.~\eqref{algorithm}.  To further demonstrate the generality of the heuristic in Eq.~\eqref{algorithm}, in this section we present results from the application of it to ASC problems, instead of the above problem-specific prescription.  Our experimental results, summarized in Fig.~\ref{fig:ASC}b-h, show that the application of anneal offsets chosen using the heuristic in Eq.~\eqref{algorithm} improves performance on ASC instances for nearly all of the problem class parameter space that was tested.

Figure~\ref{fig:ASC}b shows the success probability, $p_{\rm{success}}$, as a function of $|\delta|_{\max}$, for various problem sizes $N$ ($W_1=-1.0, W_2=-0.5, n=4$).  A performance improvement is observed for a broad range of $|\delta|_{\max}$ values, with a peak at $|\delta|_{\max}\approx0.056$, and the success probabilities dropping roughly symmetrically for smaller and greater values of $|\delta|_{\max}$.  Note that the value of $|\delta|_{\max}$ that maximizes the success probability does not change appreciably with $N$.  This is because the times at which the light and heavy sectors order depend primarily on the parameters $W_1, W_2, n$, and approximately the same value of $|\delta|_{\max}$ should synchronize the dynamics of the two kinds of sectors if these three parameters are kept fixed.  In Fig.~\ref{fig:ASC}c, we can see how the time-to-solution speed-up, TTS\textsuperscript{BL}/TTS\textsuperscript{AO}, scales with problem size both for a fixed value of $|\delta|_{\max}$ chosen based on Fig.~\ref{fig:ASC}b such that performance is improved for every $N$  ($|\delta|_{\max}=0.056$), and how it scales when at each $N$ we use the value of $|\delta|_{\max}$ that maximizes the success probability for that $N$.  In both cases, the speed-up appears to scale exponentially with $N$.  We note that $|\delta|_{\max}=0.056$ either maximizes or very nearly maximizes the success probability for every $N$ tested. 

Figure~\ref{fig:ASC}d shows $p_{\rm{success}}$ as a function of $|\delta|_{\max}$ for various values of the light coupling, $W_2$ ($W_1=-1.0, N=150, n=4$).  A performance improvement is observed for every value of $W_2$ for some $|\delta|_{\max}$.  While there is no clear relationship between the optimal choice of $|\delta|_{\max}$ and $W_2$, in Fig.~\ref{fig:ASC}e we can see that one can find a fixed value of $|\delta|_{\max}$ such that performance is either maximized or nearly maximized ($|\delta|_{\max}=0.056$) for all values of $W_2$.   Naively, one might expect $p_{\rm{BL}}$ to increase monotonically with $W_2/W_1$, for $0<W_2/W_1\leq 1$.  Intuitively, the smaller  $W_2/W_1$, the more inhomogeneous the dynamics of the light and heavy sectors, suggesting the problem might be more difficult to solve under homogeneous driving.  Indeed, this is what we see in Fig.~\ref{fig:ASC}d.  Similarly, one might intuitively expect the maximum speed-up obtained with the anneal offsets heuristic to decrease monotonically with $W_2/W_1$, for $0<W_2/W_1\leq 1$: the smaller  $W_2/W_1$, the more inhomogeneous the dynamics of the light and heavy sectors, and therefore, potentially, the more room for there is for the use of anneal offsets to provide an improvement.  While such monotonic behavior is indeed observed for $2/8\leq W_2/W_1\leq 7/8$, there is a single, stark exception to this intuition when $W_2/W_1=1/8$.  It is unclear what accounts for this exception.

Figure~\ref{fig:ASC}f shows the success probability as a function of $|\delta|_{\max}$ for various values of the sector size, $n$ ($W_1=-1.0, W_2=-0.5, N=200$) (for clarity of presentation, Fig.~\ref{fig:ASC}f shows the data for only 4 of the 18 different sector sizes tested).  A performance improvement is seen for some values of $n$, with the degree of improvement being very strongly correlated with the baseline success probability for the instance (Fig.~\ref{fig:ASC} (g and h)).  As was previously reported in Ref.~\cite{mishra2018finite}, instead of the $p_{\rm{BL}}$ dropping monotonically with $n$, instead we see in Fig.~\ref{fig:ASC}g that $p_{\rm{BL}}$ achieves a minimum for some intermediate $n^{*}$ ($n^{*}=5$), and then rises again for sector sizes larger than $n^{*}$.  In Fig.~\ref{fig:ASC}h we can see that there is a trend for the more difficult instances to benefit to a larger degree from the use of anneal offsets. 

\subsection*{Weak-Strong-Cluster (WSC) problems}

\begin{figure}[!htb]
\vspace{-7mm}
\includegraphics[width=.9\textwidth, keepaspectratio]{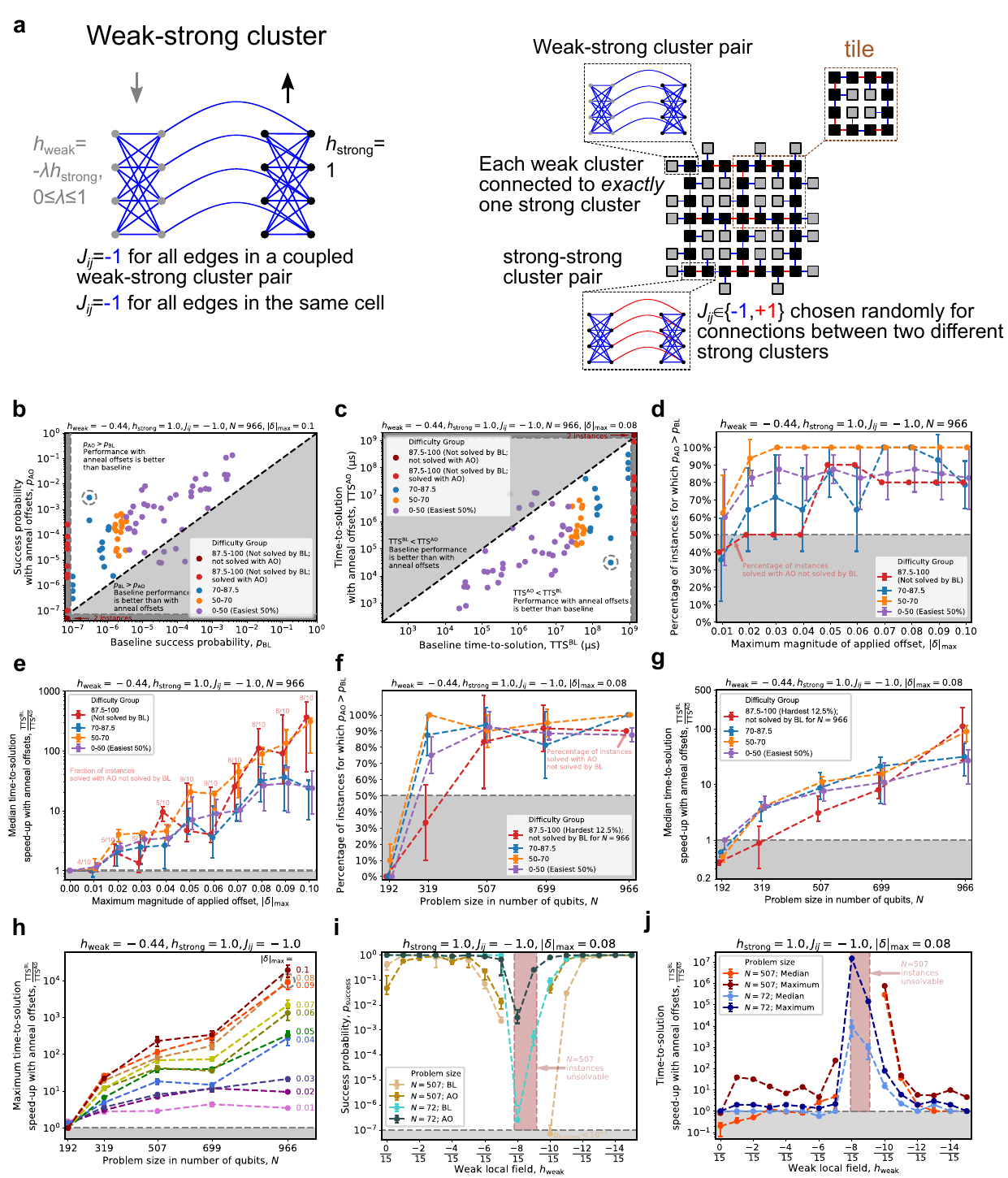}
\begin{flushleft}
\caption{}\label{fig:WSC}\footnotesize{\noindent{\bf{Fig.~\ref{fig:WSC}}}  Results for the Weak-Strong Cluster (WSC) problem class.  {\bf{a}} (Left) Weak-strong cluster pair. (Right) WSC problem example instance.  Grey blocks represent weak clusters, and black blocks represent strong clusters.  Blue lines represent ferromagnetic couplings, and red lines represent antiferromagnetic couplings.   {\bf{b}}  Instance-by-instance comparison of the observed success probability when using anneal offsets versus the observed success probability with the baseline schedule. {\bf{c}}  The corresponding times-to-solution, calculated directly from the success probabilities in {\bf{b}}.  The instance for which the maximum improvement was observed is emphasized by a grey, dashed circle.  Instances in the white zone correspond to those for which the schedule with anneal offsets resulted in better performance, whereas those in the light grey zone correspond to those for which the baseline schedule resulted in better performance.  The color indicates the relative difficulty of the instance as measured by the performance with the baseline schedule.  Note that the instances in the darker shade of red were not solved after $10^{7}$ runs neither using anneal offsets nor the baseline schedule.  Instances in the lighter shade of red were solved using the anneal offsets heuristic, but not with the baseline schedule.  {\bf{d}} Percentage of instances for which an improvement in the success probability when using anneal offsets was observed, versus the maximum magnitude of the applied offsets, $|\delta|_{\max}$. {\bf{e}} Median time-to-solution speed-up observed when using anneal offsets versus $|\delta|_{\max}$.  {\bf{f}} Percentage of instances for which an improvement in the success probability when using anneal offsets was observed, versus $N$.  {\bf{g}} Median time-to-solution speed-up observed when using anneal offsets versus $N$.  {\bf{h}} The maximum TTS speed-up observed when using anneal offsets versus $N$, for all the different values of $|\delta|_{\max}$ used.  {\bf{i}} $p_{\rm{AO}}$ and $p_{\rm{BL}}$ versus the weak local field, $h_{\text{weak}}$.  Note that $h_{\text{weak}}$ is held fixed (at $h_{\text{weak}}=-0.44$) in the rest of the figure.  {\bf{j}} The maximum and median TTS speed-ups for the instances in {\bf{i}}.}
\end{flushleft}
\end{figure}

The weak-strong cluster (WSC) class of problems was studied in the context of quantum annealing in Refs.~\cite{boixo2016computational,denchev2016computational,mandra2016strengths}.  The weak-strong cluster (WSC) problem class was designed so that multi-qubit tunneling strongly impacts the success probability.  The building block of this problem is a pair of strongly connected spins, also referred to as a pair of clusters.  One cluster is referred to as a strong cluster, and the other as a weak cluster; each cluster corresponds to a cell in the Chimera graph.  Within a weak-strong cluster pair, all the couplings are set ferromagnetically ($J_{ij}=-1$); all the local fields $h_i$ in the strong cluster are set to $h_{\rm{strong}}=-1$, and all the local fields in the weak cluster are set\clearpage to $h_{\rm{weak}}=-\lambda h_{\rm{strong}}$, for some $0<\lambda<1$ (depicted graphically in Fig.~\ref{fig:WSC}a).

For $\lambda<0.5$, the global minimum of a weak-strong cluster pair corresponds to the configuration in which all spins point in the direction of the strong local field. As explained in Ref.~\cite{denchev2016computational}, early in the anneal, however, the local-field terms dominate, so each spin orients itself along the direction of its own local field.  Later in the anneal, the coupling terms dominate, and the spins in the weak cluster must tunnel through an energy barrier to escape the local minimum into which they are lead during this initial phase of the anneal.  This problem class is interesting because it was designed to benefit from a computational strength of quantum annealing (multi-spin cotunneling \cite{boixo2016computational,denchev2016computational}), while simultaneously being difficult for classical simulated annealing to solve. This has made it a well-studied problem class for which a quantum speed-up might be obtained, although a speed-up against the best classical methods has not yet been achieved \cite{mandra2016strengths}.  Our experimental results, summarized in Fig.~\ref{fig:WSC}, show that anneal offsets chosen using the heuristic typically improve performance of the DW2000Q on the WSC class as measured by several different metrics.

Figure~5b shows a scatter plot of the observed success probability when using anneal offsets, $p_{\rm{AO}}$, versus the observed success probability with the baseline schedule, $p_{\rm{BL}}$, for 80 randomly generated instances with problem size $N=966$ spins and weak local field $h_{\rm{weak}}=-0.44$ (i.e., $\lambda=0.44$; this value of $\lambda$ is chosen because Refs.~\cite{boixo2016computational,denchev2016computational,mandra2016strengths} focused on instances with this choice of $\lambda$), using $|\delta|_{\max}=0.08$.  Figure~5c shows the corresponding times-to-solution, calculated directly from the success probabilities in Fig.~5b.  Before we continue to discuss the results in more detail, we note that 10 out of 80 instances were not solved with the baseline schedule after $10^7$ runs.  Furthermore, not all of these instances were solved with the anneal offsets heuristic; depending on the value of $|\delta|_{\max}$ used, different fractions of them were solved.  For all values of $|\delta|_{\max}$, the instances not solved using the heuristic were a strict subset of the instances not solved with the baseline schedule.  Because we were unable to obtain concrete estimates of some of the success probabilities (both $p_{\rm{AO}}$ and $p_{\rm{BL}}$), we are forced to present a somewhat tailored interpretation of the results for the corresponding instances.  In particular, for instances solved with anneal offsets but not with the baseline schedule, the times-to-solution are calculated by setting $p_{\rm{BL}}=x$, where $x$ is the smallest value such that we can conclude $P(p_{\rm{BL}}\leq x)\geq95\%$.  The degree of confidence is calculated assuming each run of the quantum annealer is a Bernoulli trial with probability of success $p_{\rm{BL}}$, and probability of failure $1-p_{\rm{BL}}$.

With this in mind, we now continue on to discuss the results. A performance improvement is observed for nearly all instances in Fig.~\ref{fig:WSC} (B and C) (with performance worsening on only 5/80 instances, all of which lie in the easiest $50\%$ of instances).  Figure~\ref{fig:WSC}D shows the percentage of instances for which using anneal offsets resulted in improved performance compared to baseline, versus $|\delta|_{\max}$.   We can see that, when $|\delta|_{\max}$ is chosen appropriately (e.g., $|\delta|_{\max}=0.08$), one can improve performance on a larger percentage of the more difficult instances, relative to the easier instances.  Similarly, we can see in Fig.~\ref{fig:WSC}E, which shows the median times-to-solution speed-ups (i.e., the median of TTS\textsuperscript{BL}/TTS\textsuperscript{AO}) versus $|\delta|_{\max}$, that one can improve performance on the more difficult instances to a greater degree, relative to the easier instances, with appropriate choice of $|\delta|_{\max}$ (e.g., $|\delta|_{\max}=0.08$).  Note that the values reported for the hardest 10/80 instances represent lower bounds for the median time-to-solution speed-up for the subset of the instances solved with the heuristic; the subset is in general different for different values of $|\delta|_{\max}$.

We now discuss how the use of the anneal offsets heuristic affects performance as a function of problem size, $N$.  Figure~\ref{fig:WSC}f shows the percentage of instances for which the use of the heuristic improved performance for various problem sizes, with $|\delta|_{\max}$ fixed at a value chosen based on Fig.~\ref{fig:WSC}d such that performance averaged over all instances is improved ($|\delta|_{\max}=0.08$).  While the behavior is different for each difficulty group, over {\it{all}} instances the overall impression is that the benefit obtained from using anneal offsets increases up to some intermediate size ($N=507$), beyond which the results remain roughly constant (with performance increasing on $\approx 90\%$ of instances).  Fig.~\ref{fig:WSC}g shows the median time-to-solution ratio for various problem sizes, with $|\delta|_{\max}$ fixed at a value chosen based on Fig.~\ref{fig:WSC}e such that performance averaged over all instances is improved ($|\delta|_{\max}=0.08$).  While the behavior again depends on the difficulty group being considered, over {\it{all}} instances  there is a trend for the benefit obtained from using anneal offsets to increase with problem size.  Note that for the largest problem size tested ($N=966$), the median speed-up reported for the hardest $12.5\%$ of instances ($\approx 160\times$) is in fact a lower bound for the actual median time-to-solution speed-up for a subset of the instances, namely, the subset of instances solved with the heuristic.  Figure~\ref{fig:WSC}h shows the maximum observed speed-up (i.e., the maximum of TTS\textsuperscript{BL}/TTS\textsuperscript{AO}) for various problem sizes.  In general, there appears to be a tendency for the maximum observed speed-up to increase with problem size; while it remains nearly constant for intermediate problem sizes (between $N=507$ by $N=699$), a positive trend resumes when including data for the largest problem size tested ($N=966$, for which a $\approx 20000 \times$ speed-up is observed).

We also study how the use of anneal offsets affects the QA performance on instances constructed with different choices of the parameter $\lambda\in[0,1]$, or, equivalently, the value of the weak local fields, $h_{\rm{weak}}$. Figure~\ref{fig:WSC}i shows the median of the observed success probabilities for 10 randomly generated instances for each of several values of $h_{\rm{weak}}$, and Fig.~\ref{fig:WSC}j shows the median and maximum time-to-solution speed-ups observed for these instances.  There appears to be a range of $h_{\rm{weak}}$ values for which there is a noticeable spike in the difficulty of the problem instances (approximately $8/15<|h_{\rm{weak}}|<9/15$ for $N=72$, and $8/15<|h_{\rm{weak}}|<10/15$ for  $N=507$).  Outside of this range, the instances in this problem class can solved with relatively high success probability when using the baseline schedule.  Inside this range, instances with a corresponding $p_{\rm{BL}}<10^{-7}$ were very common.  In fact, for $N=507$, none of the instances in this regime that we generated were solved after $10^7$ anneal runs.  In order to collect data for this parameter regime, we had to run experiments at the significantly reduced problem size of $N=72$. Notably on one of the instances the use of anneal offsets improved the success probability by $\approx 10^{7} \times$.  Consequently the time-to-solution was reduced from just over $9\times 10^8~{\rm{\mu s}}$ $(\approx 15~\rm{minutes})$ to just under $40~{\rm{\mu s}}$.  In general, we found that the use of anneal offsets improved the performance on more difficult instances to a larger degree, relative to on easier instances.

We conclude this section by noting the following. A previous benchmark study of the D-Wave 2X quantum annealer (a 1000-qubit QA) on WSC problem instances reported \cite{mandra2016strengths} that the QA was very close to achieving a quantum advantage against a battery of state-of-the-art classical solvers across all tested problem sizes. A re-evaluation using the DW2000Q with anneal offsets is an interesting prospect for future study, especially since the WSC problems benchmarked in Ref.~\cite{mandra2016strengths} were of the same subclass we show results for in Fig.~\ref{fig:WSC}~(b-h) (namely $h_\textrm{weak} = -0.44$; $h_\textrm{strong} = 1$; $J_{ij} = -1$), and in this work we have found that substantial performance improvements can be achieved on this subclass.

\subsection*{1-D Weak-Strong-Cluster Chain (1-D-WSCC) problems}

\begin{figure*}[!htb]
\caption{}\label{fig:1DWSCC}
  \includegraphics[width=0.8\textwidth, keepaspectratio]{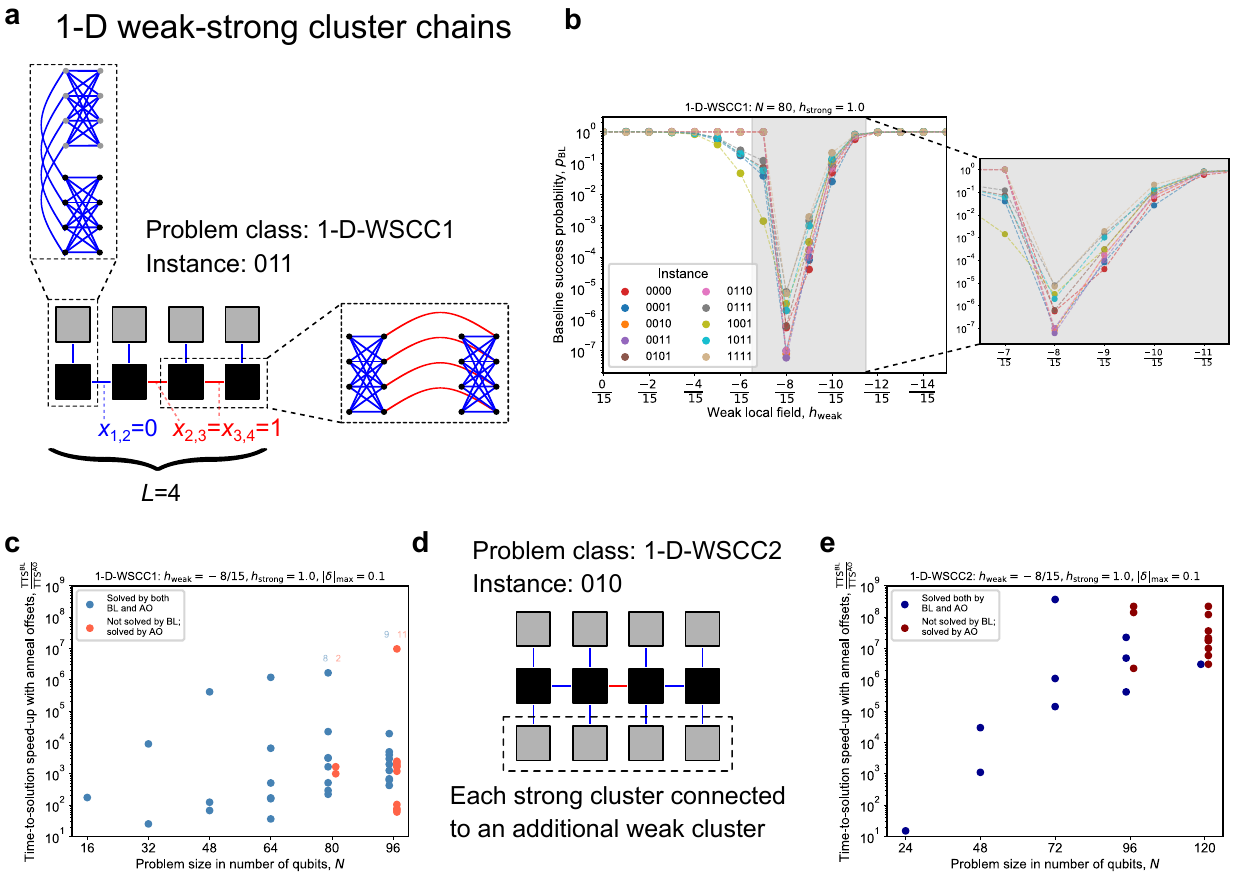}
\end{figure*}

The 1-D Weak-Strong-Cluster Chain (1-D-WSCC) problem class is a variant of the WSC problem class investigated in the previous section.  The building block of the 1-D-WSCC problem class is the weak-strong cluster (depicted graphically in Fig.~\ref{fig:WSC}a).    Adding connections between the strong clusters of a line of weak-strong clusters in such a way that the clusters form a 1-dimensional chain (depicted graphically in Fig.~\ref{fig:1DWSCC}a) yields the subclass 1-D-WSCC1 (the reason for the 1 at the end of the name will become more apparent soon).

\vspace{-9mm}
\vspace{-4mm}
  \begin{flushleft}
  \footnotesize{\noindent{\bf{Fig.~\ref{fig:1DWSCC}}}  Results for the 1-D Weak-Strong Cluster Chains problem class.  {\bf{a}} Graphical depiction of the subclass 1-D-WSCC1.  We can identify each instance with a binary string $x_{1,2}...x_{L-1,L}$, where $L$ is the length of the chain.  $x_{i,i+1}=0$ means that all connections between strong clusters $i$ and $i+1$ are set to $-1$; $x_{i,i+1}=1$ means that all connections between strong clusters $i$ and $i+1$ are set to $+1$.  Grey blocks represent weak clusters, and black blocks represent strong clusters.  Blue lines represent ferromagnetic couplings, and red lines represent antiferromagnetic couplings.    {\bf{b}} Baseline success probability, $p_{\rm{BL}}$, for various values of the weak local field, $h_{\rm{weak}}$.  Data points of the same color correspond correspond to instances with the same topology.  {\bf{c}}  Time-to-solution (TTS) speed-up with anneal offsets ($|\delta|_{\max}=0.1$) for all topologically distinct instances for various problem sizes $N$. $\rm{TTS}^{\rm{AO}}$ denotes the time-to-solution when using anneal offsets, and $\rm{TTS}^{\rm{BL}}$ denotes the time-to-solution with the baseline schedule.  The blue and orange data points have been offset for clarity, but both correspond to a problem size of either 80 or 96.  {\bf{d}} Graphical depiction of the subclass 1-D-WSCC2. {\bf{e}} Time-to-solution (TTS) speed-up with anneal offsets ($|\delta|_{\max}=0.1$) for all topologically distinct instances in the problem class 1-D-WSCC2 for various problem sizes $N$.  The dark blue and dark red data points have been offset clarity, but both correspond to a problem size of either 96 or 120.}
  \end{flushleft}
  
Figure~\ref{fig:1DWSCC}b shows a scatter plot of the observed success probability with the baseline schedule, $p_{\rm{BL}}$, for all topologically distinct instances of problem size $N=80$, for various values of the weak local field, $h_{\rm{weak}}$.  Note that there is a sharp decrease in $p_{\rm{BL}}$ approximately in the range $-7/15\geq h_{\rm{weak}}\geq-11/15$, with the minimum empirically observed at $h_{\rm{weak}}=-8/15$.  Figure~\ref{fig:1DWSCC}c shows a scatter plot of the time-to-solution ratios, $\rm{TTS}^{\rm{BL}}/\rm{TTS}^{\rm{AO}}$, for {\emph{all}} topologically distinct instances of problem size $N$, for various problem sizes $N$, with $h_{\rm{weak}}~=~-8/15$; $\rm{TTS}^{\rm{AO}}$ denotes the time-to-solution when using anneal offsets, and $\rm{TTS}^{\rm{BL}}$ denotes the time-to-solution with the baseline schedule.  A performance improvement is observed for all instances, with a tendency for the speed-ups to increase with $N$.  Whereas all instances were solved when using anneal offsets, not all instances were solved when using the baseline schedule.  For these instances, the times-to-solution are calculated in the same manner as for those instances in Fig.~
\ref{fig:WSC} that were solved with anneal offsets, but not with the baseline schedule: we set $p_{\rm{BL}}=x$, where $x$ is the smallest value such that we can conclude $P(p_{\rm{BL}}\leq x)\geq95\%$.  The degree of confidence is calculated assuming each run of the quantum annealer is a Bernoulli trial with probability of success $p_{\rm{BL}}$, and probability of failure $1-p_{\rm{BL}}$.

Figure~\ref{fig:1DWSCC}d shows a different subclass of 1-D-WSCC that can be obtained from 1-D-WSCC1 by connecting an additional weak cluster to each strong cluster.  We denote this subclass as 1-D-WSCC2; the 2 is due to the fact that each strong cluster now has 2 weak clusters connected to it, in contrast to problems in 1-D-WSCC1, which only have 1.  Figure~\ref{fig:1DWSCC}e shows a scatter plot of the time-to-solution ratios, $\rm{TTS}^{\rm{BL}}/\rm{TTS}^{\rm{AO}}$, for {\emph{all}} topologically distinct instances of various problem sizes, for $h_{\rm{weak}}=-8/15$.  As in Fig.~\ref{fig:1DWSCC}c, again we see a performance improvement for all instances, with a tendency for the speed-ups to increase with $N$.  Whereas all instances were solved when using anneal offsets, not all instances were solved when using the baseline schedule.  For these instances, the times-to-solution are calculated in the same manner as for those in Fig.~\ref{fig:1DWSCC}c.  Note that for the largest problem sizes investigated, $N=120$, the time-to-solution speed-ups are all greater than $3\times10^{6}\times$, with two speed-ups larger than $10^{8}\times$.

\section{Discussion}

The heuristic we have tested has a single parameter, $|\delta|_{\max}$, which is freely chosen. We found that different instances benefit differently for the same value of $|\delta|_{\max}$.  In general, it appears that the more difficult instances of a problem class are more likely to benefit across a wider range of $|\delta|_{\max}$ values, as well as more likely to benefit to a larger degree, relative to the easier instances in that problem class.  Of course, in practice, one does not in general know the difficulty of a particular instance \emph{a priori}. It would be helpful to develop a method capable of predicting the optimal $|\delta|_{\max}$ value to use for a particular instance based on its $h_i, J_{ij}$ values; this is an avenue for future work. For the present, our empirical results provide a guideline for choosing $|\delta|_{\max}$: if one has no other information, begin with $|\delta|_{\max} = 0.05$. This choice provides a good balance between improving the performance for difficult instances and not excessively decreasing the performance for easy instances; this was true across all broad problem classes we explored.  The scatter plots in Figures A\ref{fig:Uk_supp}, A\ref{fig:Mk_supp}, and A\ref{fig:WSC_supp} allow one to build intuition for the trade-offs obtained by various choices of $|\delta|_{\max}$. Relatdely, recall that by using the hybrid strategy one can use a higher value of $\left|\delta\right|_{\max}$ and obtain the bulk of the performance improvements from doing so while simultaneously limiting the downside for instances that aren't solved more easily using anneal offsets.

A quantitative understanding of how anneal offsets applied using the approach in this paper improves performance is currently lacking. The Alternating-Sectors-Chain problem instances are analytically tractable, which has made it possible to explore how anneal offsets change the minimum gap and the number of thermally accessible excited states, even for large problems sizes. In the Appendix we show that neither change explains the improvement in performance that we experimentally observed for this problem class. A more detailed analysis of the dynamics of the computation process might ultimately be necessary to develop a predictive model for the impact of anneal offsets; such a model could aid efforts to design optimal strategies for applying anneal offsets.

In conclusion, in this paper we demonstrated a heuristic strategy for tuning the anneal offsets on a DW2000Q quantum annealer that results in improved time-to-solution over baseline DW2000Q performance for a broad range of problems.  For the most generic problem class we investigated, the Uniform-Range-$k$-Disorder problems, one can improve the performance for up to $74\%$ of instances overall, and for $85\%$ of the hardest $10\%$ of instances, with speed-ups of up to $10^3 \times$ observed.  For more structured problem classes, like Weak-Strong-Cluster problems, one can improve the performance for up to $94\%$ of instances overall, and $100\%$ of the hardest $10\%$ of instances, with speed-ups of up to $10^8\times$ observed. Furthermore, for certain parameter regimes of WSC and ASC problems we found that the speed-ups achieved by using anneal offsets increased exponentially as a function of problem size $N$, suggesting that speed-ups orders of magnitude larger than even $10^8 \times$ may be achievable.  We anticipate that the strategy we described, and derivatives thereof, will form a useful part of the toolbox in experimental quantum annealing. The speedups we obtain for broad classes of problems naturally suggest this is a technique of potential practical relevance when attempting to optimize the performance of a quantum annealer. Furthermore, its dramatic impact on the success probabilities of Alternating-Sectors-Chain problem instances, for which we have detailed analytical results, may make this heuristic strategy a useful control knob for studies attempting to elucidate the working mechanisms of experimental quantum annealers.

{\it Note added:} A key component of the anneal-offsets heuristic presented in this paper is that a qubit's average effective field is obtained by summing over all possible neighboring spin configurations, each configuration having equal weight. This procedure can naturally be generalized to a weighted average for the effective field.  During preparation of this manuscript we became aware of related unpublished work along this line by D-Wave Systems Inc., wherein it is proposed that these weights be chosen according to the frequency with which they appear in the solutions returned by running the problem instance with the baseline schedule \cite{DWAO}.

\section{Materials and Methods}

\subsection*{D-Wave run protocol}

Each call to the DW2000Q is limited to $10^4$ annealing runs.  The success probabilities were estimated by executing up to $10^3$ calls, i.e., $10^7$ annealing runs, and stopping the calls after the instance was solved a minimum of $n_{\rm{success}}=5$ times.  For some of the more difficult problem instances, $10^7$ runs were not enough to solve the problem 5 times, but in nearly all cases it was enough to solve the problem at least once; the instances that were unsolved after $10^7$ runs are explicitly denoted in our results. A different gauge was used every $10^3$ runs.

With the exception of instances in the Alternating-Sectors-Chain (ASC) problem class, we used the Hamze-de Freitas-Selby solver (HFS) \cite{HFS} with default settings to find putative ground-state energies, and it is these energies that we use to define whether a run of the DW2000Q for a particular problem instance resulted in a success or not. For the ASC problem instances, the ground-state energy can easily be calculated exactly, so HFS was not needed to define success for any of the instances in that problem class.

\subsection*{Error analysis}

\subsubsection*{UR$k$D}

Error bars in Fig.~2 (d and f) correspond to the standard deviation when dividing the data into 4 bins.  Error bars in Fig.~2 (e, g, and i) indicate the 35th and 65th percentiles.  Finally, error bars in Fig.~2h correspond to the standard deviation across the 4 largest speed-ups observed.

\subsubsection*{ASC}
Error bars in Fig.~4 (b, d, f, and g) correspond to  $95\%$ confidence intervals.  For Fig.~4 (c, e, and h), each data point was collected 4 times.  The values presented correspond to the mean, and the error bars correspond to the standard deviation.

\subsubsection*{WSC}
Error bars in Fig.~5 (d and f) correspond to the standard deviation when dividing the data into 4 bins.  Error bars in Fig.~5 (e, g, i, and j) indicate the 35th and 65th percentiles. Finally, error bars in Fig.~5h correspond to the standard deviation across the 4 largest speed-ups observed. 

\section{Acknowledgements}
We would like to thank M. Amin, E. Hoskinson, T. Lanting, J. Dunn, and A. Mishra for useful discussions.  We would also like to thank B. Fuller and J. Brahm for a thorough reading of a draft of this paper. Finally, we gratefully acknowledge USRA and NASA for providing us with D-Wave 2000Q machine time on the system installed at NASA Ames, and in particular Davide Venturelli.

\clearpage

\vspace{5mm}
\bibliography{anneal_offsets_paper}

\begin{thebibliography}{10}

\bibitem{kadowaki1998quantum}
T.~Kadowaki and H.~Nishimori, ``Quantum annealing in the transverse {Ising}
  model,'' {\em Phys. Rev. E}, vol.~58, no.~5, p.~5355, 1998.

\bibitem{farhi2000quantum}
E.~Farhi, J.~Goldstone, S.~Gutmann, and M.~Sipser, ``Quantum computation by
  adiabatic evolution. https://arxiv.org/abs/quant-ph/0001106,'' 2000.

\bibitem{santoro2006optimization}
G.~E. Santoro and E.~Tosatti, ``Optimization using quantum mechanics: quantum
  annealing through adiabatic evolution,'' {\em Journal of Physics A:
  Mathematical and General}, vol.~39, no.~36, p.~R393, 2006.

\bibitem{johnson2011quantum}
M.~W. Johnson, M.~H.~S. Amin, S.~Gildert, T.~Lanting, F.~Hamze, N.~Dickson,
  R.~Harris, A.~J. Berkley, J.~Johansson, P.~Bunyk, E.~M. Chapple, C.~Enderud,
  J.~P. Hilton, K.~Karimi, E.~Ladizinsky, N.~Ladizinsky, T.~Oh, I.~Perminov,
  C.~Rich, M.~C. Thom, E.~Tolkacheva, C.~J.~S. Truncik, S.~Uchaikin, J.~Wang,
  B.~Wilson, and G.~Rose, ``Quantum annealing with manufactured spins,'' {\em
  Nature}, vol.~473, no.~7346, pp.~194--198, 2011.

\bibitem{ronnow2014defining}
T.~F. R{\o}nnow, Z.~Wang, J.~Job, S.~Boixo, S.~V. Isakov, D.~Wecker, J.~M.
  Martinis, D.~A. Lidar, and M.~Troyer, ``Defining and detecting quantum
  speedup,'' {\em Science}, vol.~345, no.~6195, pp.~420--424, 2014.

\bibitem{boixo2014evidence}
S.~Boixo, T.~F. R{\o}nnow, S.~V. Isakov, Z.~Wang, D.~Wecker, D.~A. Lidar, J.~M.
  Martinis, and M.~Troyer, ``Evidence for quantum annealing with more than one
  hundred qubits,'' {\em Nature Physics}, vol.~10, no.~3, pp.~218--224, 2014.

\bibitem{denchev2016computational}
V.~S. Denchev, S.~Boixo, S.~V. Isakov, N.~Ding, R.~Babbush, V.~Smelyanskiy,
  J.~Martinis, and H.~Neven, ``What is the computational value of finite-range
  tunneling?,'' {\em Phys. Rev. X}, vol.~6, no.~3, p.~031015, 2016.

\bibitem{hen2015probing}
I.~Hen, J.~Job, T.~Albash, T.~F. R{\o}nnow, M.~Troyer, and D.~A. Lidar,
  ``Probing for quantum speedup in spin-glass problems with planted
  solutions,'' {\em Phys. Rev. A}, vol.~92, no.~4, p.~042325, 2015.

\bibitem{mandra2016strengths}
S.~Mandr{\`a}, Z.~Zhu, W.~Wang, A.~Perdomo-Ortiz, and H.~G. Katzgraber,
  ``Strengths and weaknesses of weak-strong cluster problems: A detailed
  overview of state-of-the-art classical heuristics versus quantum
  approaches,'' {\em Phys. Rev. A}, vol.~94, no.~2, p.~022337, (2016).

\bibitem{albash2017evidence}
T.~Albash and D.~A. Lidar, ``Evidence for a limited quantum speedup on a
  quantum annealer. https://arxiv.org/abs/1705.07452,'' 2017.

\bibitem{mandra2017deceptive}
S.~Mandr{\`a} and H.~G. Katzgraber, ``A deceptive step towards quantum speedup
  detection,'' {\em {Quant. Sci. Technol.}}, vol.~3, p.~04LT01, 2018.

\bibitem{rams2016inhomogeneous}
M.~M. Rams, M.~Mohseni, and A.~del Campo, ``Inhomogeneous quasi-adiabatic
  driving of quantum critical dynamics in weakly disordered spin chains,'' {\em
  New Journal of Physics}, vol.~18, no.~12, p.~123034, 2016.

\bibitem{mohseni2018engineering}
M.~Mohseni, J.~Strumpfer, and M.~M. Rams, ``Engineering non-equilibrium quantum
  phase transitions via causally gapped {Hamiltonians}.
  https://arxiv.org/abs/1804.11037,'' 2018.

\bibitem{campo2013universal}
A.~del Campo, T.~W.~B. Kibble, and W.~H. Zurek, ``Causality and non-equilibrium
  second-order phase transitions in inhomogeneous systems,'' {\em Journal of
  Physics: Condensed Matter}, vol.~25, no.~40, p.~404210, 2013.

\bibitem{nishimori2017exponential}
H.~Nishimori and K.~Takada, ``Exponential enhancement of the efficiency of
  quantum annealing by non-stoquastic {Hamiltonians},'' {\em Frontiers in ICT},
  vol.~4, p.~2, 2017.

\bibitem{susa2017relation}
Y.~Susa, J.~F. Jadebeck, and H.~Nishimori, ``Relation between quantum
  fluctuations and the performance enhancement of quantum annealing in a
  nonstoquastic {Hamiltonian},'' {\em Phys. Rev. A}, vol.~95, no.~4, p.~042321,
  2017.

\bibitem{hormozi2017nonstoquastic}
L.~Hormozi, E.~W. Brown, G.~Carleo, and M.~Troyer, ``Nonstoquastic
  {Hamiltonians} and quantum annealing of an {Ising} spin glass,'' {\em Phys.
  Rev. B}, vol.~95, no.~18, p.~184416, 2017.

\bibitem{vinci2017non}
W.~Vinci and D.~A. Lidar, ``Non-stoquastic {Hamiltonians} in quantum annealing
  via geometric phases,'' {\em {npj} Quantum Information}, vol.~3, no.~1,
  p.~38, 2017.

\bibitem{susa2018exponential}
Y.~Susa, Y.~Yamashiro, M.~Yamamoto, and H.~Nishimori, ``Exponential speedup of
  quantum annealing by inhomogeneous driving of the transverse field,'' {\em
  Journal of the Physical Society of Japan}, vol.~87, no.~2, p.~023002, 2018.

\bibitem{gomez2018universal}
F.~G{\'o}mez-Ruiz and A.~del Campo, ``Universal dynamics of inhomogeneous
  quantum phase transitions: suppressing defect formation,'' {\em arXiv
  preprint arXiv:1805.00525}, 2018.

\bibitem{venturelli2018reverse}
D.~Venturelli and A.~Kondratyev, ``Reverse quantum annealing approach to
  portfolio optimization problems,'' {\em arXiv preprint arXiv:1810.08584},
  2018.

\bibitem{mott2017solving}
A.~Mott, J.~Job, J.-R. Vlimant, D.~Lidar, and M.~Spiropulu, ``Solving a {Higgs}
  optimization problem with quantum annealing for machine learning,'' {\em
  Nature}, vol.~550, no.~7676, pp.~375--379, 2017.

\bibitem{lokhov2018optimal}
A.~Y. Lokhov, M.~Vuffray, S.~Misra, and M.~Chertkov, ``Optimal structure and
  parameter learning of {Ising} models,'' {\em Science Advances}, vol.~4,
  no.~3, p.~e1700791, 2018.

\bibitem{li2018quantum}
R.~Y. Li, R.~Di~Felice, R.~Rohs, and D.~A. Lidar, ``Quantum annealing versus
  classical machine learning applied to a simplified computational biology
  problem,'' {\em {npj} Quantum Information}, vol.~4, no.~1, p.~14, 2018.

\bibitem{andriyash2016factoring}
{D-Wave Systems Inc.} {\em D-Wave Technical Report Series \emph{14-1002A-B}},
  Burnaby, BC, 2016.

\bibitem{lanting2017experimental}
T.~Lanting, A.~D. King, B.~Evert, and E.~Hoskinson, ``Experimental
  demonstration of perturbative anticrossing mitigation using nonuniform driver
  {Hamiltonians},'' {\em Phys. Rev. A}, vol.~96, no.~4, p.~042322, 2017.

\bibitem{DW2000Q}
{D-Wave Systems Inc.}, ``{The D-Wave 2000Q\textsuperscript{TM} System}.
  https://www.dwavesys.com/d-wave-two-system.''
\newblock Last accessed on 2018-05-23.

\bibitem{boixo2016computational}
S.~Boixo, V.~N. Smelyanskiy, A.~Shabani, S.~V. Isakov, M.~Dykman, V.~S.
  Denchev, M.~H. Amin, A.~Y. Smirnov, M.~Mohseni, and H.~Neven, ``Computational
  multiqubit tunnelling in programmable quantum annealers,'' {\em Nature
  Communications}, vol.~7, p.~10327, 2016.

\bibitem{mishra2018finite}
A.~Mishra, T.~Albash, and D.~A. Lidar, ``Finite temperature quantum annealing
  solving exponentially small gap problem with non-monotonic success
  probability,'' {\em Nature communications}, vol.~9, no.~1, p.~2917, 2018.

\bibitem{dickson2013thermally}
N.~G. Dickson, M.~W. Johnson, M.~H. Amin, R.~Harris, F.~Altomare, A.~J.
  Berkley, P.~Bunyk, J.~Cai, E.~M. Chapple, P.~Chavez, F.~Cioata, T.~Cirip,
  P.~deBuen, M.~Drew-Brook, C.~Enderud, S.~Gildert, F.~Hamze, J.~P. Hilton,
  E.~Hoskinson, K.~Karimi, E.~Ladizinsky, N.~Ladizinsky, T.~Lanting, T.~Mahon,
  R.~Neufeld, T.~Oh, I.~Perminov, C.~Petroff, A.~Przybysz, C.~Rich, P.~Spear,
  A.~Tcaciuc, M.~C. Thom, E.~Tolkacheva, S.~Uchaikin, J.~Wang, A.~B. Wilson,
  Z.~Merali, and G.~Rose, ``Thermally assisted quantum annealing of a 16-qubit
  problem,'' {\em Nature Communications}, vol.~4, p.~1903, 2013.

\bibitem{mohseni2018constructing}
M.~Mohseni and H.~Neven.
\newblock US Patent 9,934,468 (2018).

\bibitem{marshall2018power}
J.~Marshall, D.~Venturelli, I.~Hen, and E.~G. Rieffel, ``The power of pausing:
  advancing understanding of thermalization in experimental quantum
  annealers,'' {\em arXiv preprint arXiv:1810.05881}, 2018.

\bibitem{test1}
M.~Amin. private communication, 2017.

\bibitem{zhu2016best}
Z.~Zhu, A.~J. Ochoa, S.~Schnabel, F.~Hamze, and H.~G. Katzgraber, ``Best-case
  performance of quantum annealers on native spin-glass benchmarks: How chaos
  can affect success probabilities,'' {\em Phys. Rev. A}, vol.~93, no.~1,
  p.~012317, 2016.

\bibitem{reichardt2004quantum}
B.~W. Reichardt, ``The quantum adiabatic optimization algorithm and local
  minima,'' in {\em Proceedings of the Thirty-sixth Annual ACM Symposium on
  Theory of Computing}, pp.~502--510.
\newblock Chicago, Illinois, June 13-15, 2003 (ACM, New York, NY, 2004).

\bibitem{sachdev2011quantum}
S.~Sachdev, {\em Quantum Phase Transitions}.
\newblock Cambridge University Press, ed. 2, 2011.
\newblock [second edition].

\bibitem{DWAO}
{D-Wave Systems Inc.}, ``{D-Wave Online Learning}.
  https://cloud.dwavesys.com/learning/user/
  [user]/notebooks/2000q/04-feature-examples/anneal-offset-module3.ipynb.''
\newblock Last accessed on 2018-06-20.

\bibitem{HFS}
{Alex Selby}, ``{Hamze-de Freitas-Selby solver (HFS)}.\\
  https://github.com/lanl-ansi/hfs-algorithm.''
\newblock Last accessed on 2018-07-04.

\bibitem{Note1}
It is by no means a given that the results of a closed-system analysis will
  carry over into the open-system case, but the goal in this section is to
  develop some intuition rather than establish any rigorous results.

\end{thebibliography}
\bibliographystyle{ieeetr}
\onecolumngrid
\appendix
\setcounter{figure}{0}
\section{Generalizations of the heuristic}

In the heuristic used in the paper, the key quantity in determining the anneal offsets to be applied on a particular qubit is its effective field averaged {\emph{uniformly}} over all possible neighboring spin values.  However, this can be generalized to arbitrary {\emph{weighted}} averages in a straightforward way by introducing nonnegative weights, which sum to 1, for each qubit and each configuration of its neighboring spins. In other words, for each $i\in\{1,\dots,N\}$ and each $s_{j_1},\dots,s_{j_{N_i}}\in\{-1,+1\}$, we introduce a $p^i_{s_{j_1},\dots,s_{j_{N_i}}}$ such that 
\begin{equation}
\sum_{s_{j_1},\dots,s_{j_{N_i}}\in \{-1,+1\}} p^i_{s_{j_1},\dots,s_{j_{N_i}}}=1,
\end{equation}
and look the corresponding weighted average
\begin{equation}\label{generalization}
\overline{|\mathcal{F}_i|}^{\{p^i\}}\coloneqq \sum_{s_{j_1},
\dots,s_{j_{N_i}}\in \{-1,+1\}} p^i_{s_{i_1},\dots,s_{j_{N_i}}}|\mathcal{F}_i(s_{j_1},\dots,s_{j_{N_i}})|.
\end{equation}

In Ref.~\cite{DWAO}, it is proposed that these weights be chosen according to the frequency with which the corresponding configuration appears in the solutions returned by running the problem instance with the baseline schedule several times.  They test their method on one problem instance of size $N=72$, and show that their strategy indeed improves performance.  We note that random sampling approaches such as the one in Ref.~\cite{DWAO} present a modification to the heuristic in Eq.~(6) that could enable the heuristic to be tractable even when the hardware graph has a large maximum degree. Benchmarking such strategies on a broad set of problems is an interesting future direction one could pursue. 

\section{The challenge of finding optimal anneal offsets}

The task of finding optimal offsets is in general difficult.  The following intuitive argument partially explains why this is the case.

For simplicity, let us consider this problem in the framework of closed-system adiabatic quantum computing\footnote{It is by no means a given that the results of a closed-system analysis will carry over into the open-system case, but the goal in this section is to develop some intuition rather than establish any rigorous results.}.  Let $|\psi(s)\rangle$
denote the quantum state of the $N$-qubit processor at normalized time $s\in[0,1]$. Let $U_{\bm{\delta}}(s)$
be the unitary evolution operator induced by the Hamiltonian $H_{\bm{\delta}}$ of Eq.~\eqref{AO QA} (i.e., $U_{\bm{\delta}}(s)|\psi(0)\rangle=|\psi(s)\rangle$).  Let $X=[\delta_1^{\min},\delta_1^{\max}]\times\dots\times[\delta_N^{\min},\delta_N^{\max}]\subset\mathbb{R}^{N}$ denote the set of offset values the hardware can physically implement.  In principle, the optimal choice of offsets, which we denote $\bm{\delta}^*$, is then given by
\begin{equation}\label{Optimal Offsets}
\bm{\delta}^*=\argmax_{\bm{\delta}\in X} p_{0}(\bm{\delta}) = \argmax_{\bm{\delta}\in X} \sum_{d=1}^{d_{0}} |\langle\psi_{d}^{0}|U_{\bm{\delta}}(1)|\psi(0)\rangle|^{2},
\end{equation}
where $|\psi^0_1\rangle,...,|\psi^0_{d_{0}}\rangle$
are the $d_{0}$-degenerate ground states of the problem Hamiltonian, and $p_{0}(\bm{\delta})$ denotes the probability of success.  This is a high-dimensional, non-linear, constrained optimization problem.  Furthermore, Eq.~\eqref{Optimal Offsets} suggests the success probability depends on the ground states $|\psi_{1}^{0}\rangle,...,|\psi_{d_{0}}^{0}\rangle$
of the problem.  In other words, in order to determine the choice of offsets that will maximize the probability of solving the problem, one has to know the answers in advance.  This defeats, at least in practice, the purpose of finding optimal offsets in the first place.  Indeed, Eq.~\eqref{Optimal Offsets} is not practically solvable in full generality by any single, currently known computational method.  Therefore, if one is interested in treating the problem analytically, one likely has to assume enough knowledge of the problem so that Eq.~\eqref{Optimal Offsets} can be simplified enough that an analytic treatment becomes possible.  On the other hand, assuming such knowledge a priori reduces the scope of applicability of the treatment in practice.

\section{Additional Results}

In this section we present additional experimental results for some the problem classes we tested, and well as results for a problem class not treated discussed in the main text.  While the summary statistics presented in the main text have the advantage of being concise, we find that there is nevertheless a lot of additional, more subtle information that they do not quite capture.

\subsection{Uniform-Range-$k$-Disorder (UR$k$D) problems}

\begin{figure}[!htb]
\vspace{-10mm}
\caption{}\label{fig:Uk_supp}
\includegraphics[width=0.95\textwidth, keepaspectratio]{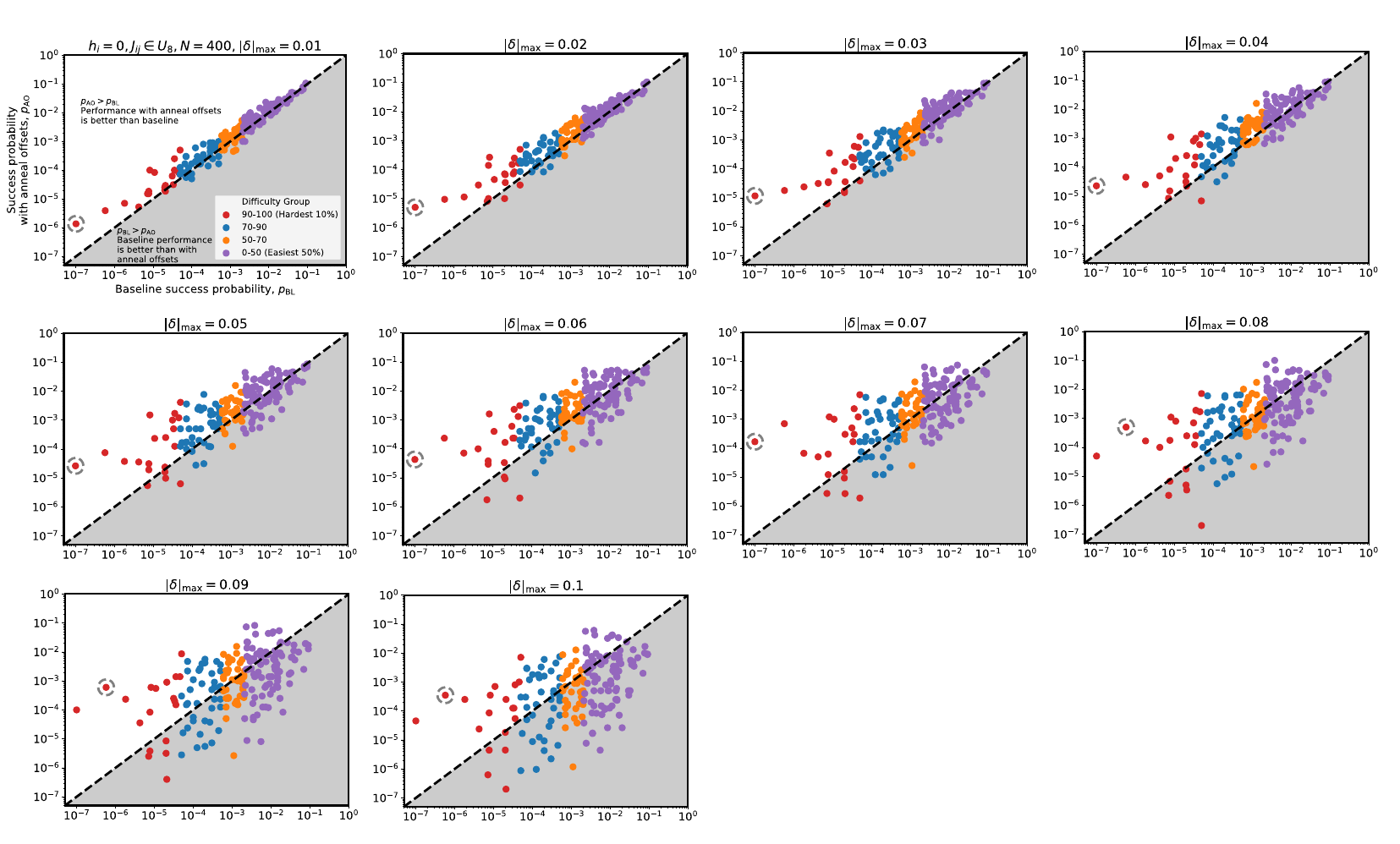}
\begin{flushleft}
\footnotesize{\noindent{\bf{Fig.~A\ref{fig:Uk_supp}.  Instance-by-instance comparison of the observed success probability when using anneal offsets, $\bm{p_{\rm{AO}}}$, versus the observed success probability with the baseline schedule, $\bm{p_{\rm{BL}}}$, for the UR8D problem class for every value of $\bm{|\delta|_{\max}}$ tested; problem size $\bm{N=400}$.}}  In general, the variance in the ratio $p_{\rm{BL}}/p_{\rm{AO}}$ increases with $|\delta|_{\max}$.}
\end{flushleft}
\end{figure}

Figure~A\ref{fig:Uk_supp} shows an instance-by-instance comparison of the observed success probability when using anneal offsets, $p_{\rm{AO}}$, versus the observed success probability with the baseline schedule, $p_{\rm{BL}}$, for the U8RD problem class of problem size $N=400$, for every value of $|\delta|_{\max}$ tested.  We can see that there is a considerable amount of variance both in the baseline success probabilities, as well as the success probabilities when using anneal offsets.  Furthermore, there is a significant amount of variance in the degree to which anneal offsets either improves or worsens the success probability, with the variance becoming increasingly pronounced with larger $|\delta|_{\max}$.  This high-variance nature of the data makes it difficult to summarize in just a few summary statistics.

\clearpage

\subsection*{4-Modal-Range-$k$-Disorder (4MR$k$D) problems}

\begin{figure}[!htb]
\caption{}
\label{fig:Mk}
\includegraphics[width=0.9\textwidth, keepaspectratio]{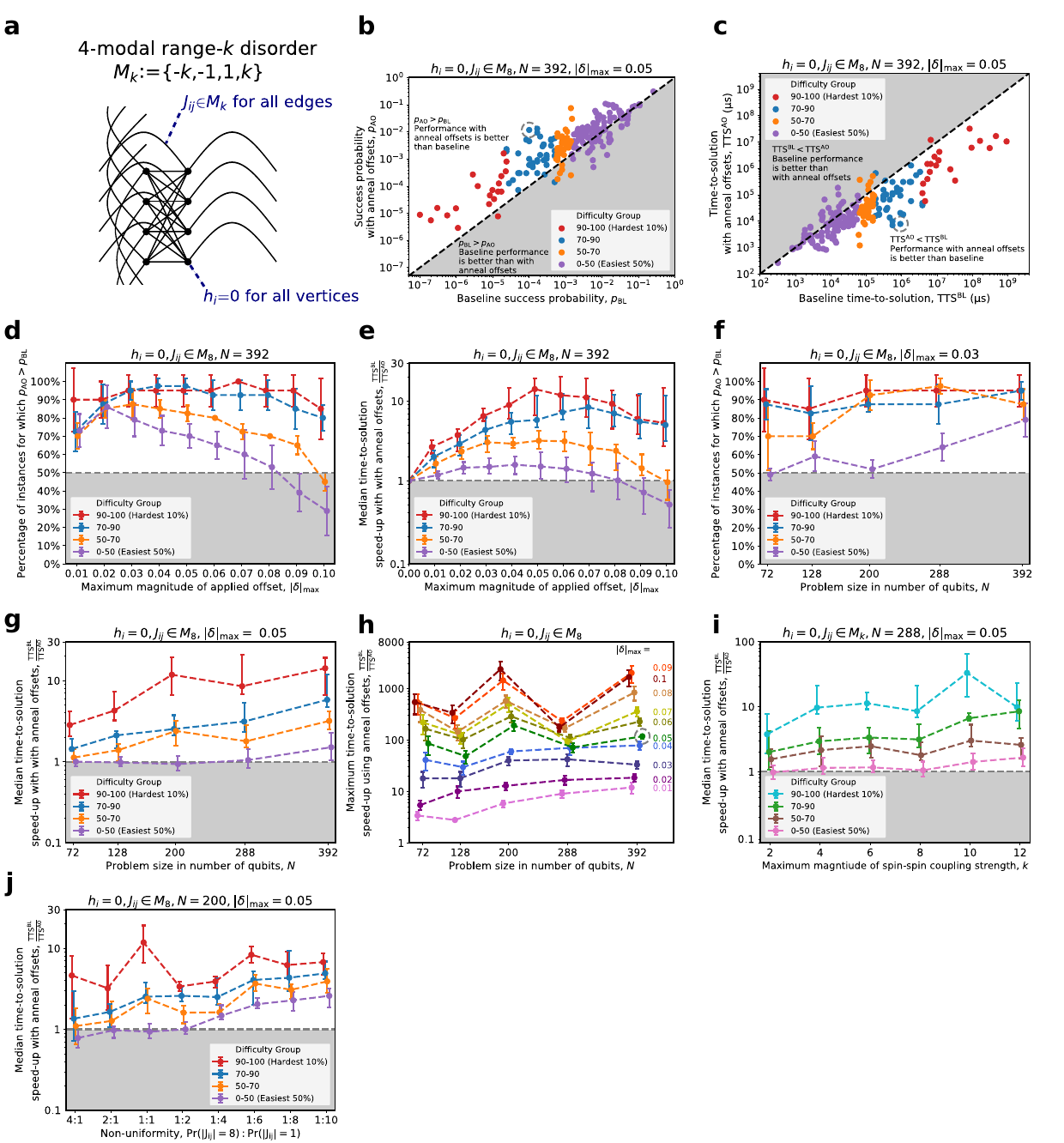}
\begin{flushleft}
\footnotesize{\noindent{\bf{Fig.~A\ref{fig:Mk}.  Results for the 4-Modal Range-$\bm{k}$ Disorder (4MR$k$D) problem class.}} {\bf{a}} A zoom-in on one of the Chimera cells in a 4MR$k$D disorder problem instance.  The coupler values $J_{ij}$ are chosen at random, with uniform probability, from the set $M_k\coloneqq\{-k,-1,1,k\}$.  Note that $0\notin M_k$.  The fields $h_v$ are all set to 0 (i.e, $h_v=0$ for all $v\in V$).  {\bf{b}}  Instance-by-instance comparison of the observed success probability when using anneal offsets, $p_{\rm{AO}}$, versus the observed success probability with the baseline schedule, $p_{\rm{BL}}$. {\bf{c}}  The corresponding times-to-solution, calculated directly from the success probabilities in {\bf{b}}; $\rm{TTS}^{\rm{AO}}$ denotes the time-to-solution when using anneal offsets, and $\rm{TTS}^{\rm{BL}}$ denotes the time-to-solution with the baseline schedule.  The instance for which the maximum improvement was observed is emphasized by a grey, dashed circle.  Instances in the white zone correspond to those for which the schedule with anneal offsets resulted in better performance, whereas those in the grey zone correspond to those for which the baseline schedule resulted in better performance.  The color indicates the relative difficulty of the instance as measured by the performance with the baseline schedule.  {\bf{d}} Percentage of instances for which an improvement in the success probability when using anneal offsets was observed, versus the maximum magnitude of the applied offsets, $|\delta|_{\max}$. {\bf{e}} Median TTS speed-up observed when using anneal offsets versus $|\delta|_{\max}$.  {\bf{f}} Percentage of instances for which an improvement in the success probability when using anneal offsets was observed, versus $N$.  {\bf{g}} Median TTS speed-up observed when using anneal offsets versus $N$.  {\bf{h}} Median TTS speed-up observed when using anneal offsets for different probability distributions with which the $J_{ij}$ are drawn from the set $M_k$.  {\bf{i}} The maximum TTS speed-up observed when using anneal offsets versus $N$, for all the different values of $|\delta|_{\max}$ used.  {\bf{j}} Median TTS speed-up observed when using anneal offsets versus the spin-spin coupling range, $k$.  Note that $k$ is held fixed (at $k=8$) in the rest of the figure.}
\end{flushleft}
\end{figure}

The 4-Modal-Range-$k$-Disorder (4MR$k$D) class of problems is defined (Fig.~A\ref{fig:Mk}a) on the Chimera graph as those for which $h_i=0$ for all $i\in\{1,\dots,N\}$, and $J_{ij}$ is chosen at random, with uniform probability, from the four discrete values in the set $M_k\coloneqq\{-k,-1,1,k\}$.  As with the UR$k$D class, we generated random instances of problems in the 4MR$k$D class for various problem sizes $N$ and coupling ranges $k$, and measured the success probability of finding a ground state for each instance, with and without the use of anneal offsets. Our experimental results, summarized in Fig.~A\ref{fig:Mk}, show that, again, anneal offsets chosen using the heuristic in Eq.~\eqref{algorithm} typically result in improved performance.

Figure~A\ref{fig:Mk}b shows a scatter plot of the observed success probability when using anneal offsets, $p_{\rm{AO}}$, versus the observed success probability with the baseline schedule, $p_{\rm{BL}}$, for 200 randomly generated instances of problem size $N=392$ and maximum magnitude of spin-spin coupling $k=8$, using a maximum magnitude of applied offset $|\delta|_{\max}=0.05$.  Figure~A\ref{fig:Mk}c shows the corresponding times-to-solution, calculated directly from the success probabilities in Fig.~A\ref{fig:Mk}b; $\rm{TTS}^{\rm{AO}}$ denotes the time-to-solution when using anneal offsets, and $\rm{TTS}^{\rm{BL}}$ denotes the time-to-solution with the baseline schedule.  As with the UR$k$D problem class, we note that it is again the case that the more difficult instances (for the baseline solver, as defined by $p_{\rm{BL}}$) are both more likely to benefit from the use of anneal offsets, as well as more likely to benefit to a larger degree (relative to the easier instances in this problem class).  One can see this more clearly in Fig.~A\ref{fig:Mk}d,e, which show the percentage of instances for which using anneal offsets resulted in improved performance compared to baseline and the median time-to-solution ratio (i.e., the median of $\rm{TTS}^{\rm{BL}}/\rm{TTS}^{\rm{AO}}$), respectively, both as a function of $|\delta|_{\max}$.

\begin{figure}[!h]
\includegraphics[width=1\textwidth, keepaspectratio]{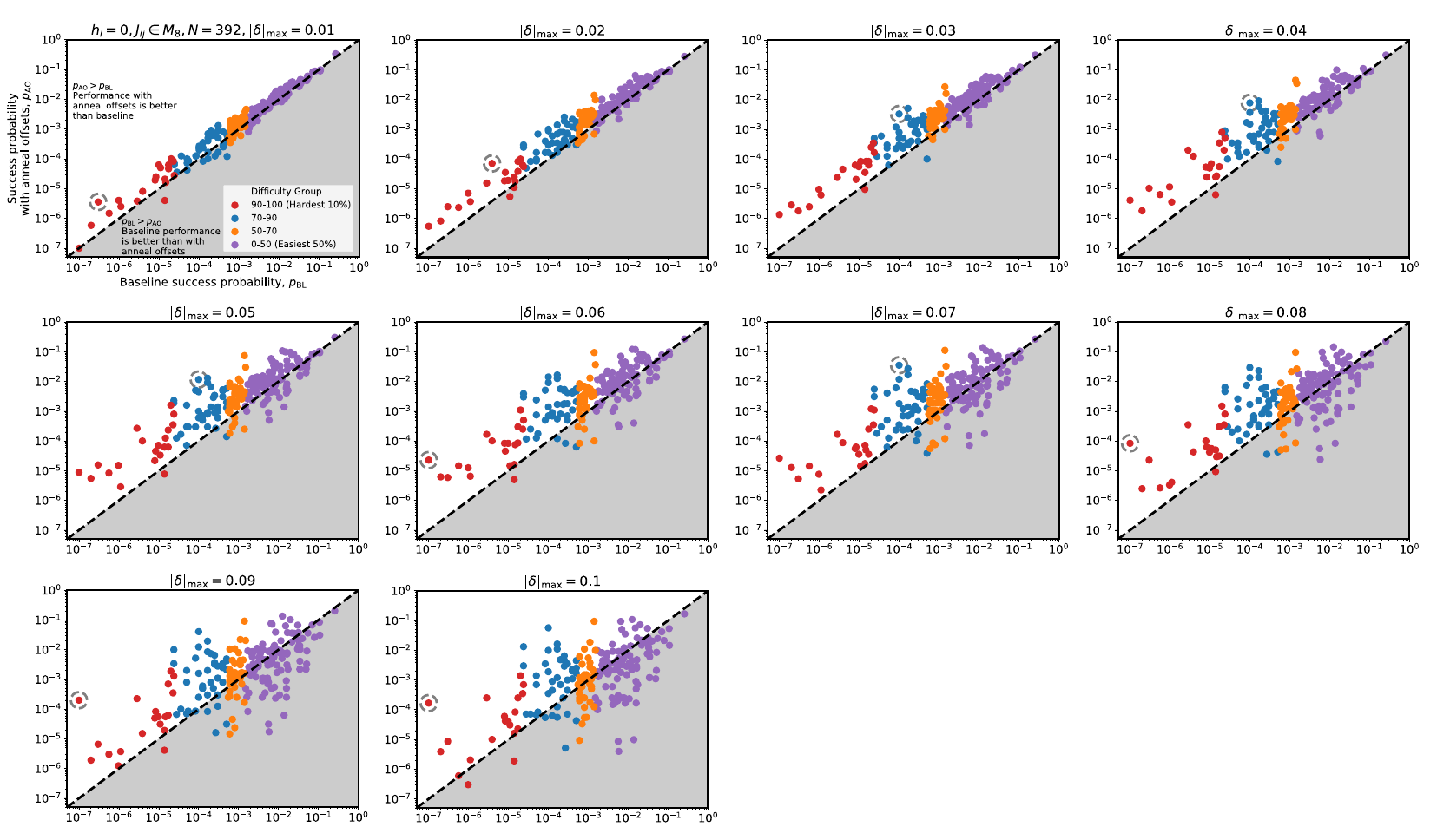}
\caption{}\label{fig:Mk_supp}
\begin{flushleft}
\footnotesize{\noindent{\bf{Fig.~A\ref{fig:Mk_supp}.  Instance-by-instance comparison of the observed success probability when using anneal offsets, $\bm{p_{\rm{AO}}}$, versus the observed success probability with the baseline schedule, $\bm{p_{\rm{BL}}}$, for the 4MR8D problem class for every value of $\bm{|\delta|_{\max}}$ tested; problem size $\bm{N=392}$.}}  In general, the variance in the ratio $p_{\rm{BL}}/p_{\rm{AO}}$ increases with $|\delta|_{\max}$.}
\end{flushleft}
\end{figure}

It is again difficult to pick a single value of $|\delta|_{\max}$ that results in the ``best'' performance for the 4MR$k$D problem class, since determining the best choice of $|\delta|_{\max}$ depends on the performance metric used to measure its optimality, and because different instances will be affected differently for the same value of $|\delta|_{\max}$ (Fig.~A\ref{fig:Mk}d,e).  The primary trade-offs to be balanced are again that smaller values of $|\delta|_{\max}$ will generally result in increased performance over a larger percentage of the instances compared to larger values of $|\delta|_{\max}$; however, smaller values of $|\delta|_{\max}$ will generally result in smaller median time-to-solution speed-ups compared to larger values of $|\delta|_{\max}$ (see Fig.~A\ref{fig:Mk_supp} for more details).

We now discuss how the performance of the heuristic depends on problem size.  Figure~A\ref{fig:Mk}f shows the percentage of instances for which the heuristic improved performance for various problem sizes, $N$, with $|\delta|_{\max}$ fixed at a value chosen based on Fig.~A\ref{fig:Mk}d such that performance averaged over all instances is improved ($|\delta|_{\max}=0.03$).  While the metric as a function of $N$ for each difficulty group is different,  the overall impression is that as $N$ increases, the benefit obtained from using anneal offsets over baseline is either constant or increasing.  Furthermore, across all problem sizes there is a tendency for performance to be improved on a larger percentage of the more difficult instances, relative to the easier instances.  Fig.~A\ref{fig:Mk}g shows the median time-to-solution ratio for various problem sizes, with $|\delta|_{\max}$ fixed at a value chosen based on Fig.~A\ref{fig:Mk}e such that performance averaged over all instances is improved ($|\delta|_{\max}=0.05$).  While the behavior is again dependent on the difficulty group, in general there appears to be tendency for the median time-to-solution to increase slightly with problem size, with the trend slightly more pronounced than for the UR$k$D problem class.  Figure~A\ref{fig:Mk}h shows the maximum observed speed-up (i.e., the maximum of $\rm{TTS}^{\rm{BL}}/\rm{TTS}^{\rm{AO}}$) for various problem sizes, for all values of $|\delta|_{\max}$ tested.  In general, the maximum speed-up appears to be roughly constant with problem size.  The largest speed-up observed across all instances was just over $2000\times$ ($N=200, |\delta|_{\max}=0.1$).  Furthermore, we note that there is a tendency for the higher values of $|\delta|_{\max}$ to be more likely to result in the maximum speed-up, compared to the smaller values of $|\delta|_{\max}$.

The results discussed thus far have concerned the case when $k=8$, where $k$ is the maximum magnitude of the spin-spin coupling values; additionally, we have focused on the case where the edge weights $J_{ij}$ are drawn with {\it{uniform}} probability from the set $M_k$.  We now discuss the performance of the heuristic when these features of the problem class are modified.  Still drawing the $J_{ij}$ uniformly from $M_k$, the results appear to be generally consistent across a broad range of different $k$ values (Fig.~A\ref{fig:Mk}i).  Figure~A\ref{fig:Mk}j shows how the median of time-to-solution ratio changes when we change the probability distribution with which the $J_{ij}$ are drawn (keeping $k=8$ fixed).  For the easier $90\%$ of instances, there appears to be a tendency for the median time-to-solution ratio to increase slightly with ${\rm{Pr}}(|J_{ij}|=1)/{\rm{Pr}}(|J_{ij}|=8)$, meanwhile for the hardest $10\%$ of instances the ratio is roughly constant.

Figure~A\ref{fig:Mk_supp} shows an instance-by-instance comparison of the observed success probability when using anneal offsets, $p_{\rm{AO}}$, versus the observed success probability with the baseline schedule, $p_{\rm{BL}}$, for the 4MR8D problem class of problem size $N=392$, for every value of $|\delta|_{\max}$ tested.  We can see a very similar situation here as with the results in Fig.~A\ref{fig:Uk_supp}, discussed in the previous paragraph: the high-variance nature of the data makes it difficult to succinctly summarize.

\clearpage

\subsection*{Alternating-Sectors-Chain (ASC) problems}

\begin{figure}[h!]
\includegraphics[width=1\linewidth]{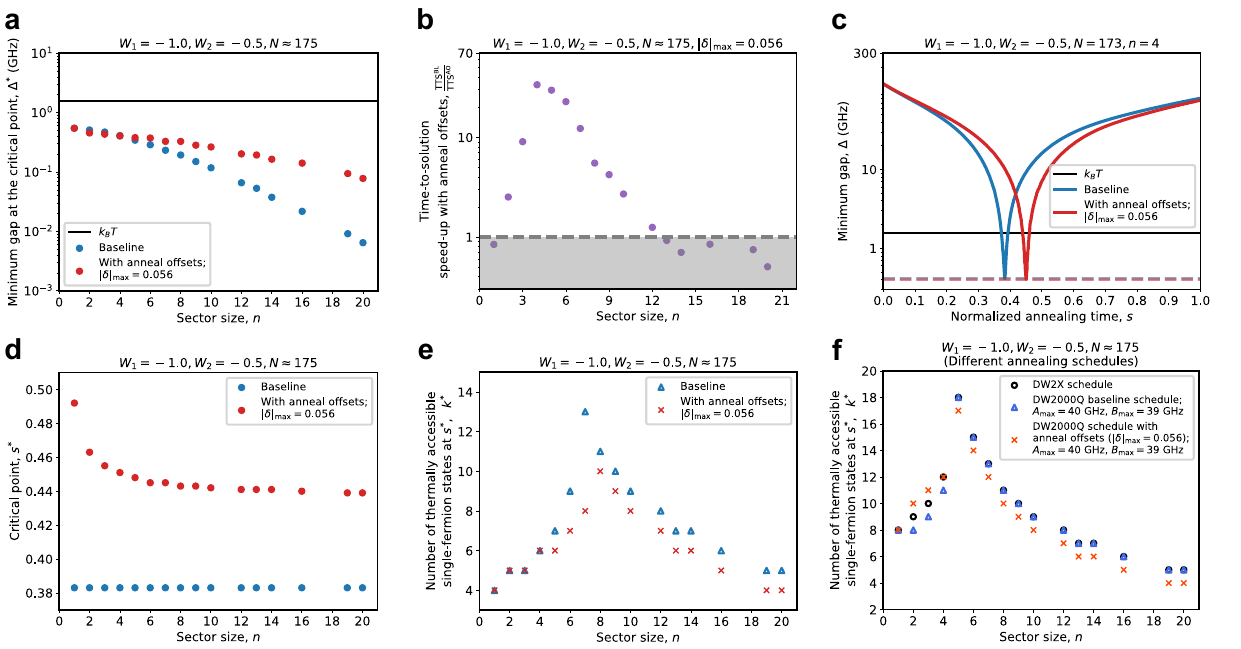}
\caption{}\label{fig:ASC_supp}
\begin{flushleft}
\footnotesize{\noindent{\bf{Fig.~A\ref{fig:ASC_supp}. Additional results for the ASC problem class with heavy coupling $\bm{W_1=1.0}$, light coupling $\bm{W_2=0.5}$, and problem size $\bm{N\approx 175}$.}}
{\bf{a}} Minimum gap at the critical point, $\Delta^{*}$, versus sector size $n$.  The solid black line indicates the energy scale set by the operating temperature.  {\bf{b}}  Observed TTS speed-up using anneal offsets compared to baseline versus $n$.  {\bf{c}} Minimum gap, $\Delta$, versus normalized annealing time, $s$, for sector size $n=4$, and problem size $N=173$.  The solid black line indicates the energy scale set by the operating temperature.  {\bf{d}}  Critical point, $s^*$, versus $n$.  {\bf{e}} Number of thermally accessible single-fermion states at the critical point, $k^*$, versus $n$.  {\bf{f}}  Comparison of $k^*$ versus $n$ using the annealing schedules of the D-Wave 2X processor in \cite{mishra2018finite}, and using the annealing schedules of the quantum annealer used in this study with the energy scales rescaled to match the energy scales used the the aforementioned D-Wave 2X.}
\end{flushleft}
\end{figure}

Figure~A\ref{fig:ASC_supp} shows additional results for the alternating sectors chain problem class of problem size $N\approx175$.  In Fig.~A\ref{fig:ASC_supp}a we can see that the minimum gap at the critical point when using the anneal offsets heuristic, which we'll denote $\Delta^{*}_{\rm{AO}}$, is larger than the baseline minimum gap at the critical point, which we'll denote $\Delta^{*}_{\rm{BL}}$,  for sector sizes $n\geq5$ (we use $h=1$ units throughout).  While both $\Delta^{*}_{\rm{AO}}$ and $\Delta^{*}_{\rm{BL}}$ decrease monotonically with $n$, $\Delta^{*}_{\rm{AO}}$ appears to decreases more slowly compared to $\Delta^{*}_{\rm{BL}}$.  One might naively expect from Fig.~A\ref{fig:ASC_supp}a that the largest speed-up observed when using anneal offsets would occur for $n=20$, where $\Delta^{*}_{\rm{AO}}/\Delta^{*}_{\rm{BL}}$ is maximized, but we can see in Fig.~A\ref{fig:ASC_supp}b that this is not the case.  Instead, the speed-up increases with sector size for $1\leq n\leq 4$, and then generally decreases for $n>4$ (with very minor exceptions at $n=16, 19$).  Indeed, there does not appear to exist any clear correlation between the ratio $\Delta^{*}_{\rm{AO}}/\Delta^{*}_{\rm{BL}}$ and the observed speed-ups.  Note that the energy scale set by the operating temperature is much larger than both $\Delta^{*}_{\rm{AO}}$ and $\Delta^{*}_{\rm{BL}}$ for every sector size.

In Fig.~A\ref{fig:ASC_supp}c we can see how the minimum gap when using anneal offsets, which we'll denote $\Delta_{\rm{AO}}$, differs from the baseline minimum gap, which we'll denote $\Delta_{\rm{BL}}$, as a function of normalized annealing time $s\in[0,1]$, when the sector size $n=4$, which is the value of $n$ at which the greatest time-to-solution speed-up is observed.  Given the substantial difference in the times-to-solution, it is somewhat surprising that the only notable difference between $\Delta_{\rm{AO}}$ and $\Delta_{\rm{BL}}$ is the shift in the critical points.  A similar shift is observed for all sector sizes (Fig.~A\ref{fig:ASC_supp}d), independent of the time-to-solution ratio observed for that value of the sector size.

In Ref.~\cite{mishra2018finite}, it is argued that a key quantity in predicting the success probability of the quantum annealer on this problem class is the number of single-fermion states that lie below the energy scale set by the temperature at the critical point, which we'll denote $k^{*}$.  In general, in Ref.~\cite{mishra2018finite} it is shown that the success probability decreases when $k^{*}$ increases.  It is therefore interesting to check what effect, if any, anneal offsets has on this quantity; one could conjecture that the speed-ups observed when using anneal offsets are a consequence of anneal offsets decreasing $k^{*}$.  In Fig.~A\ref{fig:ASC_supp}e we indeed see that $k^{*}_{\rm{AO}}\leq k^{*}_{\rm{BL}}$ for all sector sizes tested; here $k^{*}_{\rm{AO}}$ is the number of single-fermion states that lie below the energy scale set by the temperature at the critical point using the anneal offsets heuristic, and $k^{*}_{\rm{BL}}$ is the same quantity but using the baseline schedule.  Interestingly, however, that there is no clear correlation between observed speed-up and the ratio $k^{*}_{\rm{AO}}/k^{*}_{\rm{BL}}$ .  For example, the greatest speed-up is observed for sector size $n=4$, but $k^{*}_{\rm{AO}}=k^{*}_{\rm{BL}}$ for this sector size.

It is interesting to note that the values of $k^{*}_{\rm{BL}}$ (Fig.~A\ref{fig:ASC_supp}e)  differ considerably from those in \cite{mishra2018finite}.  This can be attributed in large part to the fact that the annealing schedules of the quantum annealer used in this study are different from the annealing schedules of the quantum annealer in \cite{mishra2018finite}.  In fact, by rescaling the energy (in simulation) of the annealing schedules used in this study to match the energy scales of \cite{mishra2018finite}, we get exact agreement (Fig.~A\ref{fig:ASC_supp}f) for nearly all sector sizes (with only a minimal disagreement for sector sizes $n\in\{2,3,4\}$).  It is still the case at these different energy scales that there is no clear correlation between the time-to-solution ratio and the ratio $k'^{*}_{\rm{AO}}/k'^{*}_{\rm{BL}}$; here $k'^{*}_{\rm{AO}}$ is the number of single-fermion states that lie below the energy scale set by the temperature at the critical point using the anneal offsets heuristic on a DW2000Q with the energy rescaled to match the energy of the quantum annealer used in \cite{mishra2018finite}, and $k'^{*}_{\rm{BL}}$ denotes the analogous quantity with homogeneous annealing schedules.  For example, while the greatest speed-up is observed for sector size $n=4$, $k'^{*}_{\rm{AO}}>k'^{*}_{\rm{BL}}$ for this sector size..  Similarly, $k'^{*}_{\rm{AO}}(n)<k'^{*}_{\rm{BL}}(n)$ for $n>13$, even though the performance of the quantum annealer with the anneal offsets heuristic is in fact \emph{worse} than with the baseline schedule for those sector sizes.

\clearpage

\subsection*{Weak-Strong-Cluster (WSC) problems}

\begin{figure}[!h]
\includegraphics[width=1\textwidth, keepaspectratio]{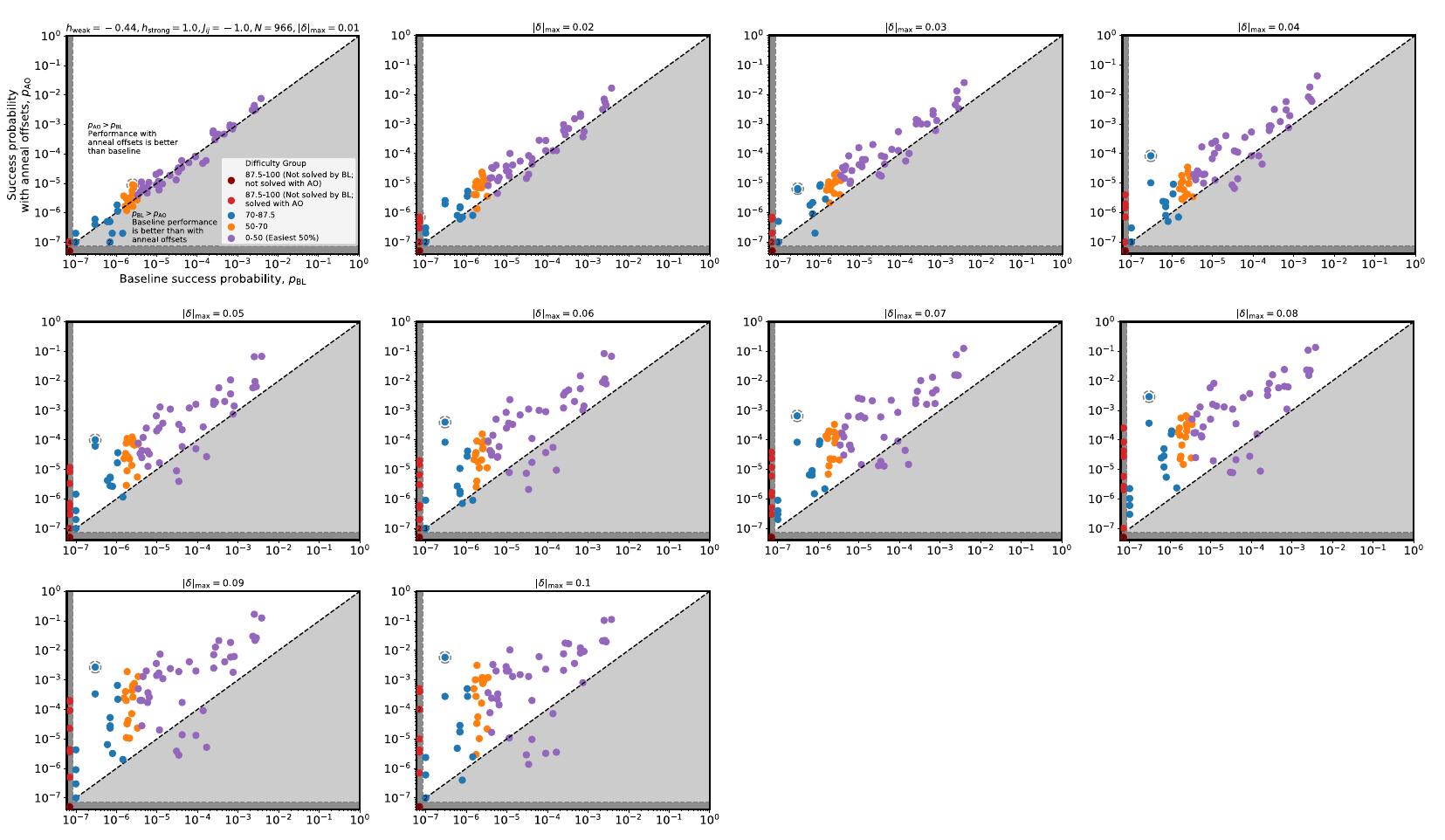}
\caption{}\label{fig:WSC_supp}
\begin{flushleft}
\footnotesize{\noindent{\bf{Fig.~A\ref{fig:WSC_supp}.  Instance-by-instance comparison of the observed success probability when using anneal offsets, $\bm{p_{\rm{AO}}}$, versus the observed success probability with the baseline schedule, $\bm{p_{\rm{BL}}}$, for the WSC problem class for every value of $\bm{|\delta|_{\max}}$ tested; problem size $\bm{N=966}$.}}  In general, the variance in the ratio $p_{\rm{BL}}/p_{\rm{AO}}$ increases with $|\delta|_{\max}$.  Interestingly, in contrast to the UR8D (Fig.~A\ref{fig:Uk_supp}) and 4MR8D ((Fig.~A\ref{fig:Mk_supp}) problem classes, the increased variance is primarily a consequence of instances benefiting to a larger degree, as opposed to a mix of some instances improving to a larger degree, and others being negatively impacted to a larger degree. Data points with numbers inscribed indicate overlapping data points.}
\end{flushleft}
\end{figure}

Figure~A\ref{fig:WSC_supp} shows an instance-by-instance comparison of the observed success probability when using anneal offsets, $p_{\rm{AO}}$, versus the observed success probability with the baseline schedule, $p_{\rm{BL}}$, for the WSC problem class of problem size $N=966$ and weak local fields $h_{\rm{weak}}=-0.44$, for every value of $|\delta|_{\max}$ tested.  Compared to both the UR8D problem instances and the 4MR8D disorder problem instances of Fig.~A\ref{fig:Uk_supp} and Fig.~A\ref{fig:Mk_supp} respectively, we can see that the WSC problem instances in A\ref{fig:WSC_supp} are more robust to anneal offsets: they benefit both to a larger degree and across a wider-range of $|\delta|_{\max}$ values.  Note that 10/80 instances were not solved with the baseline schedule after $10^{7}$ runs.  Depending on the value of $|\delta|_{\max}$, a different fraction of these instances were solved using anneal offsets, which allows one to derive at least a lower bound on the improvement when using anneal offsets.  For all values of $|\delta|_{\max}$, the instances that were not solved with the anneal offsets heuristic were a strict subset of the instances not solved with the baseline schedule (i.e., we did not observe any instance solved with the baseline schedule that was not also solved using the anneal offsets heuristic, independent of $|\delta|_{\max}$).  For these particularly difficult instances, perhaps it would be interesting in future work to perform more runs to obtain more concrete estimates on the success probabilities.  It is unclear how many runs this would require, or if it would be feasible.

\newpage

\section{The annealing schedules and the working graph of the quantum processor used in this study}

\begin{figure}[!h]
\includegraphics[width=\textwidth]{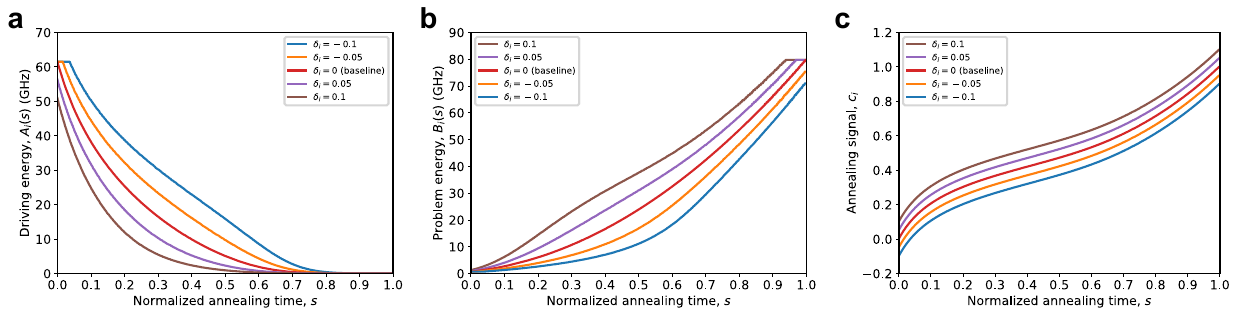}
\caption{}\label{fig:schedules}
\begin{flushleft}
\footnotesize{\noindent{\bf{Fig.~A\ref{fig:schedules}. Annealing schedules of the D-Wave 2000Q quantum annealer used in this study for different values of anneal offsets.}}  Data for the baseline schedule obtained from John Dunn of D-Wave Systems Inc.}
\end{flushleft}
\end{figure}

\begin{figure}[h!]
\includegraphics[width=\linewidth,keepaspectratio]{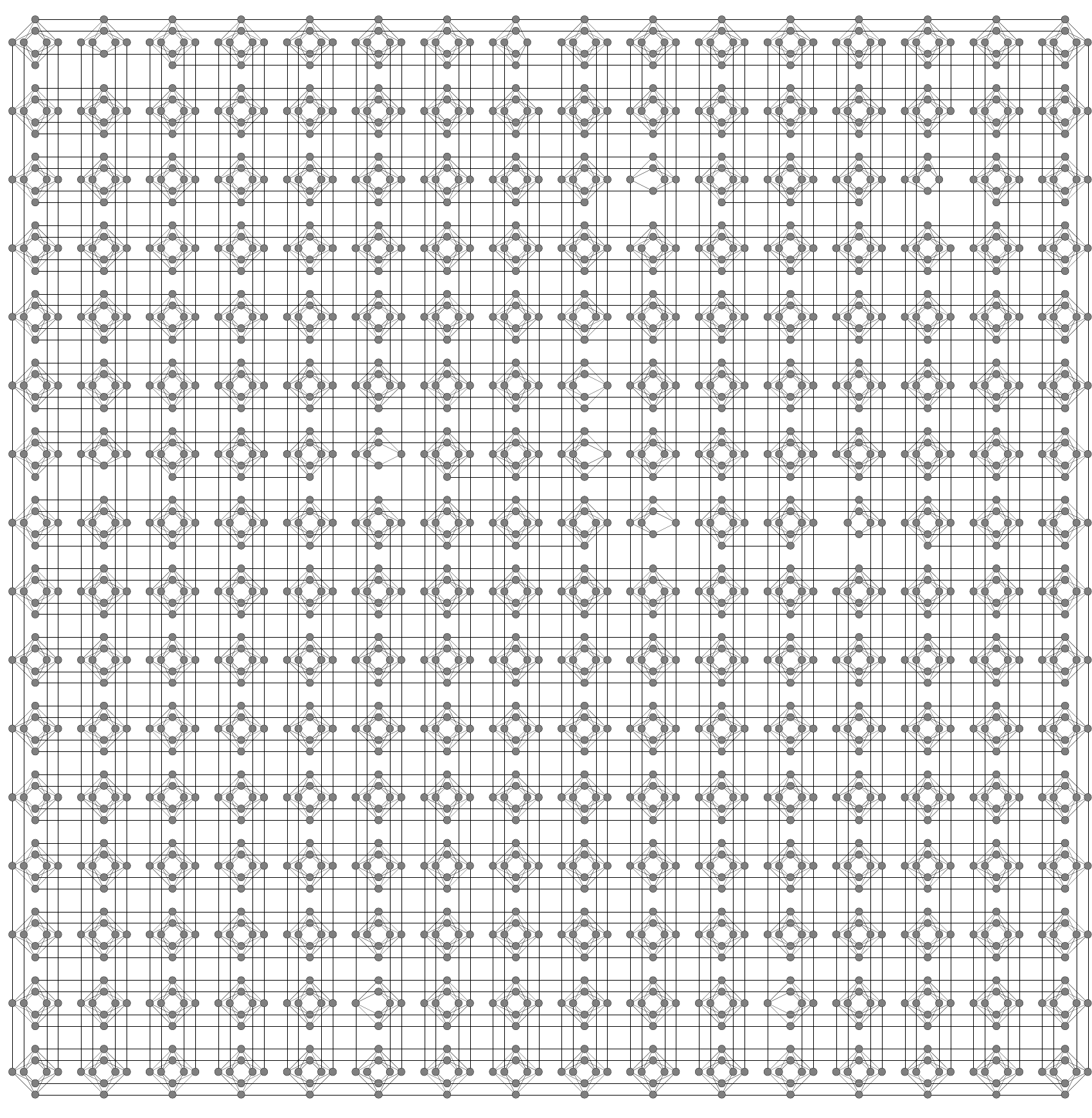}
\caption{}\label{fig:Chimera}
\begin{flushleft}
\footnotesize{\noindent{\bf{Fig.~A\ref{fig:Chimera}. Working graph of the D-Wave 2000Q quantum annealer used in this study.}}  Obtained from, and printed with permission from John Dunn of D-Wave Systems Inc.}
\end{flushleft}
\end{figure}

\end{document}